%% file: main.tex
\documentclass[preprint,1p,times,onecolumn,rgb,hyperref,table,sort&compress]{elsarticle}

\journal{Information and Software Technology}

\usepackage{packages}
\AtBeginDocument{%
  \providecommand\BibTeX{{%
    \normalfont B\kern-0.5em{\scshape i\kern-0.25em b}\kern-0.8em\TeX}}}

\begin{document}
\begin{frontmatter}

\title{Classification and Challenges of Non-Functional Requirements in ML-Enabled Systems: A Systematic Literature Review}

\author{Vincenzo De Martino}
\ead{vdemartino@unisa.it}

\author{Fabio Palomba}
\ead{fpalomba@unisa.it}

\address{Software Engineering (SeSa) Lab - Department of Computer Science, University of Salerno (Italy)}

\thispagestyle{plain}
\pagestyle{plain}



\begin{abstract}
\textbf{Context:} Machine learning (ML) is nowadays so pervasive and diffused that virtually no application can avoid its use. Nonetheless, its enormous potential is often tempered by the need to manage non-functional requirements and navigate pressing, contrasting trade-offs.
\textbf{Objective:} In this respect, we notice the lack of a comprehensive synthesis of the non-functional requirements affecting ML-enabled systems, other than the major challenges faced to deal with them. Such a synthesis may not only provide a comprehensive summary of the state of the art, but also drive further research on the analysis, management, and optimization of non-functional requirements of ML-intensive systems.
\textbf{Method:} In this paper, we propose a systematic literature review targeting two key aspects such as (1) the classification of the non-functional requirements investigated so far, and (2) the challenges to be faced when developing models in ML-enabled systems.
Through the combination of well-established guidelines for conducting systematic literature reviews and additional search criteria, we survey a total amount of 69 research articles.
\textbf{Results:} Our findings report that current research identified 30 different non-functional requirements, which can be grouped into six main classes.
We also compiled a catalog of more than 23 software engineering challenges, based on which further research should consider the nonfunctional requirements of machine learning-enabled systems.
\textbf{Conclusion:} We conclude our work by distilling implications and a future outlook on the topic.
\end{abstract}

\begin{keyword}
Software Engineering for Artificial Intelligence; Non-Functional Requirements; Systematic Literature Reviews.
\end{keyword}
\end{frontmatter}



\input{sections/1_Introduction}

\input{sections/2_Motivation}

\input{sections/3_Methodology}

\input{sections/4_Results}

\input{sections/5_Discussions}

\input{sections/6_Threat}

\input{sections/7_Conclusions}

\section*{Credits}
\textbf{Vincenzo De Martino}: Formal analysis, Investigation, Data Curation, Validation, Writing - Original Draft, Visualization.
\textbf{Fabio Palomba}: Supervision, Validation, Writing - Review \& Editing.

\section*{Conflict of interest}
The authors declare that they have no conflict of interest.

\section*{Data Availability}
The data collected in the context of this systematic literature review, along with the scripts used to analyze and generate data, charts, and plots discussed when addressing our research goals, are publicly available at:  \cite{DeMartino2023} 

\section*{Acknowledgement}
This work has been partially supported by the European Union - NextGenerationEU through the Italian Ministry of University and Research, Projects PRIN 2022 "QualAI: Continuous Quality Improvement of AI-based Systems" (grant n. 2022B3BP5S , CUP: H53D23003510006) and PRIN 2022 PNRR "FRINGE: context-aware FaiRness engineerING in complex software systEms" (grant n. P2022553SL, CUP: D53D23017340001). The opinions presented in this article solely belong to the author(s) and do not necessarily reflect those of the European Union or The European Research Executive Agency. The European Union and the granting authority cannot be held accountable for these views.

\bibliographystyle{ACM-Reference-Format}
\bibliography{bib/bibliography}

\bibliographystyleS{unsrt}
\bibliographyS{bib/formal_literature}

\end{document}

%% file: sections/1_Introduction.tex
\section{Introduction}
\label{sec:introduction}
Machine learning (ML) is now, more than ever, being used in theory, experiment, and simulation \cite{ourmazd2020science,harvard}. On the one hand, companies and individuals increasingly rely on the outcome of machine learning models to make informed decisions \cite{zhou2018human} or automate tasks that would take substantial human workload~\cite{rech2004artificial}. On the other hand, machine learning-intensive systems, i.e., systems that embed machine learning solutions, have been recently deployed in multiple domains, with some recent applications showing highly efficient and accurate performance~\cite{olson2011algorithm,miller2015can}. As such, the pervasiveness of machine learning-intensive systems is expected to increase further in the coming years in multiple domains \cite{ravi2016deep,zhang2019deep}. 

Nonetheless, such a pervasiveness is constantly threatened by multiple concerns, which are often not related to the specific features made available to users, but to non-functional attributes \cite{hochstein2023don,mitNews}. In particular, a non-functional attribute is defined as a condition that specifies a criterion that may be used to judge the operation of a system rather than specific behaviors \cite{bruegge2009object}. In the context of machine learning-enabled systems, non-function attributes may affect the overall level of reliability, trustworthiness, and sustainability of these systems  \cite{lwakatare2020large,habibullah2021non,horkoff2019non}. It is therefore not surprising that the software engineering research community---and specifically the software engineering for artificial intelligence (SE4AI) research branch---has been investing notable efforts in understanding non-functional requirements of machine learning-intensive systems, other than proposing methods and instruments to support practitioners when dealing with them \cite{habibullah2022non,binkhonain2019review,khan2022handling}. This effort is also stimulated by government and funding agencies, which are more and more willing to invest in the matter, e.g., the European Union has recently approved the so-called \textsl{Artificial Intelligence Act},\footnote{The European Union Artificial Intelligence Act: \url{https://artificialintelligenceact.eu}.} which aims at promoting research on themes connected to the improvement of non-functional attributes of artificial intelligence-based software systems. 

Recent advances in the field of SE4AI contributed to the development of a consistent body of knowledge with respect to the management of multiple non-functional requirements, including fairness \cite{brun2018software,galhotra2017fairness,zhang2021ignorance}, security \cite{hains2018towards,lo2021systematic}, privacy \cite{liu2021machine,siavvas2022technical}, and more \cite{kumeno2019sofware}. While recognizing the relevant advances made over the last years, our research identifies a notable key limitation.

\steattentionbox{\faWarning \hspace{0.05cm} 
Despite the current, extensive body of knowledge produced by the SE4AI research community with respect to the management of non-functional requirements of ML-intensive systems, there is still a lack of a comprehensive, systematic synthesis of the current knowledge on the \textbf{non-functional attributes affecting machine learning-enabled systems} and the \textbf{challenges faced when dealing with them}.}

An improved understanding of these aspects may have critical implications for researchers and practitioners. First, researchers might learn more about the current state of the art, possibly identifying neglected research angles that would be worth further investigating. At the same time, practitioners may have a comprehensive overview of the instruments that researchers have been providing to support the analysis and optimization of non-functional requirements, possibly accelerating the technological transfer of academic prototypes to industry.

In this paper, we conduct a systematic literature review (SLR) on non-functional requirements of machine learning-intensive software systems. Our work follows well-established guidelines \cite{kitchenham,10.1145/2601248.2601268} and additional search criteria based on seed set identification \cite{mourao2020performance} to comprehensively synthesize existing research. 
From an initial set composed of over 2,500 hits, and after applying multiple snowballing rounds and additional data collection procedures, we ended up analyzing more than 81,000 research results. Through the application of exclusion/inclusion criteria and a rigorous quality assessment, we finally selected 69 papers.
In addition, we provide a novel catalog composed of more than 23 software engineering challenges to deal with them proposed by researchers to support practitioners when optimizing non-functional requirements throughout the software lifecycle. We conclude the paper by elaborating on the implications of our results, along with the actionable items that readers of our work may (re-)use to analyze further the problem of non-functional requirements in machine learning-intensive software systems. 

\medskip
\noindent \textbf{Structure of the paper.} Section \ref{sec:motivation} analyzes the related work. Section \ref{sec:methodology} describes the research goals and the methods applied to address them, while Section \ref{sec:results} discusses the results achieved. The implications and actionable insights are elaborated in Section \ref{sec:discussion}, with the limitations reported in Section \ref{sec:threat}. Finally, Section \ref{sec:conclusion} concludes the paper and outlines our future research agenda.

%% file: sections/2_Motivation.tex
\begin{table*}[h]
    \centering
    \caption{Comparison with the Closest Related Work.} 
    \label{tab:related_differences}
    \rowcolors{1}{gray!15}{white}
    \resizebox{1\linewidth}{!}{
    
    \begin{tabular}{|p{0.25\linewidth} | p{0.5\linewidth} | p{0.6\linewidth}|}
    \rowcolor{black} 
    \textcolor{white}{\textbf{Related Work}} & \textcolor{white}{\textbf{Main Focus}} & \textcolor{white}{\textbf{Commonalities and differences}}\\
    
    Martinez-Fernandez et al.~\cite{martinez2022software} &
    A systematic mapping review targeting all previous research on software engineering for artificial intelligence. The authors surveyed 248 studies, classifying the available research according to the SWEBOK areas \cite{bourque2014swebok}. &
    \begin{minipage}[t]{\hsize}
        \vspace*{-2.5ex}
        \begin{itemize}[nosep,leftmargin=*]
            \raggedright
            \item Similar research approach to the search, yet with some differences in terms of seed set identification;
            
            \item A broader analysis of current research, with no specific focus on non-functional requirements;
    
            \item While the work includes part of the research papers we analyze, it does not provide insights into the challenges with non-functional requirements of machine learning-intensive systems. 
        \end{itemize}%
        \vspace{0.2cm}
    \end{minipage}%
    \\
    \hline
    Habibullah et al.~\cite{habibullah2022non} & A non-systematic analysis of non-functional requirements of machine learning-intensive systems. The authors focused on (i) the identification of clusters of non-functional requirements, (ii) the estimation of the amount of relevant studies for a subset of non-functional requirements, and (iii) the definition of the scope of non-functional requirements.
    &
    \begin{minipage}[t]{\hsize}
        \vspace*{-2.5ex}
        \begin{itemize}[nosep,leftmargin=*]
            \raggedright 
        \item We approach the literature search with a systematic approach, hence making a more comprehensive analysis;

        \item We did not limit ourselves to the estimation of the current interest in the matter but provided a systematic investigation;

        \item We let additional include challenges with non-functional requirements of machine learning-intensive systems.
        \end{itemize}%
    \end{minipage}%
    \\
    \hline
    Horkoff~\cite{horkoff2019non} &
    An experience report on the challenges that the Requirements Engineering (RE) research community is called to face when addressing non-functional attributes of machine learning-intensive systems. &
    \begin{minipage}[t]{\hsize}
        \vspace*{-2.5ex}
        \begin{itemize}[nosep,leftmargin=*]
            \raggedright 
        \item We approach the literature search with a systematic approach, hence making a more comprehensive analysis;

        \item We let additional factors arise, including a comprehensive set of non-functional requirements for machine learning-intensive systems, other than practices and methods to deal with them;

        \item We let the challenges to deal with non-functional requirements emerge from the scientific literature on the matter, hence approaching the research differently.
        \end{itemize}%
        \vspace{0.2cm}
    \end{minipage}%
    \\
    \hline
    Habibullah and Horkoff~\cite{habibullah2021non} &
    An industrial, interview-based study aiming at assessing (i) the identification and measurement of non-functional requirements, (ii) the importance of non-functional requirements in the industry, and (iii) the challenges associated with the identified non-functional requirements.
    &
    \begin{minipage}[t]{\hsize}
        \vspace*{-2.5ex}
        \begin{itemize}[nosep,leftmargin=*]
            \raggedright
        \item We let our findings emerge from the scientific literature on the matter, hence approaching the research differently;

        \item We let additional including a comprehensive set of non-functional requirements for machine learning-intensive systems;

        \item We did not limit ourselves to the observations made by the industry practitioners but provided a systematic investigation of the current relevance of non-functional requirements.
    
        \end{itemize}%
        \vspace{0.1cm}
    \end{minipage}%
    \\
    \hline
    Ahmad et al.~\cite{ahmad2023requirements} & A systematic mapping study on the requirements specification and modeling approaches and tools.
    &
    \begin{minipage}[t]{\hsize}
        \vspace*{-2.5ex}
        \begin{itemize}[nosep,leftmargin=*]
            \raggedright 
        \item The work targets requirements engineering from the perspective of specification and modeling, while ours focuses on non-functional attributes;

        \item The research method is similar, yet we perform a systematic literature review rather than a mapping study, hence being more restrictive in terms of quality assessment of the primary studies;
        \end{itemize}%
    \end{minipage}%
    \\
    \hline
    \end{tabular}
    }
\end{table*}

\section{Related Work and Motivation}
\label{sec:motivation}
To the best of our knowledge, no systematic literature review has been conducted with the aim of classifying non-functional requirements of machine learning-enabled systems and summarizing the challenges to deal with them. At the same time, it is important to point out that some secondary studies recently attempted to (1) synthesize the research on software engineering for artificial intelligence \cite{martinez2022software}, (2) explore, in a preliminary fashion, the relevance and research interest around non-functional requirements of machine learning-intensive systems \cite{habibullah2022non}, and (3) summarize some of the key academic and industrial challenges when dealing with non-functional requirements in industry \cite{horkoff2019non,habibullah2021non}. These works are clearly the most closely related to ours. This section describes the major differences and limitations of current literature that motivated our research. A summary is provided in Table \ref{tab:related_differences}. 

First and foremost, Martinez-Fernandez et al. \cite{martinez2022software} conducted a systematic mapping study of the research on software engineering for artificial intelligence. The main goal of the work was to provide a comprehensive schema representing the elements composing the field of SE4AI, from requirements engineering to verification and validation. In other terms, the systematic mapping study had a pretty broad objective and aimed at covering all the research on the matter. As such, there are multiple differences for our study. While our scope is limited to non-functional requirements, we aim to address the matter in a more detailed fashion by letting emerge a complete set of non-functional requirements discussed in the literature, other than the challenges associated to their management. Secondly, ours is a systematic literature review rather than a mapping study: as such, there are intrinsic, methodical differences in the search process conducted and in the criteria used to select the relevant pieces of research. Third, we additionally tuned the search process to produce an extensive set of seed papers, as further discussed in Section \ref{sec:methodology}---hence attempting to strengthen the completeness of the search process.

Habibullah et al. \cite{habibullah2022non} recently investigated the topic of non-functional requirements of machine learning-enabled systems under three perspectives such as (1) the clustering of non-functional attributes based on shared characteristics; (2) the estimation of the number of relevant studies that investigated aspects connected to non-functional attributes; and (3) the definition of the scope of non-functional requirements. Habibullah et al. \cite{habibullah2022non} shared the same overall objective of our paper, i.e., an improved understanding of non-functional requirements of machine learning-intensive systems. Nonetheless, we aimed to conduct a comprehensive, systematic literature review. At the same time, we focused on a larger set of research angles, including an investigation of the challenges to deal with non-functional requirements. 

Horkoff \cite{horkoff2019non} discussed the challenges that the requirements engineering research community would be called to face to address themes connected to non-functional requirements of machine learning-intensive systems. The author identified several challenges based on the extensive experience accumulated in the industry over the years. With respect to this paper, our systematic literature review aims to collect comprehensive pieces of information coming from the scientific community to provide insights into how researchers have addressed the problem so far. At the same time, it is worth saying that we could corroborate the relevance of some of the challenges identified by Horkoff \cite{horkoff2019non} through the evidence coming from our work. 

Habibullah and Horkoff \cite{habibullah2021non} conducted an interview-based study with ten machine learning engineers to elicit the practices used to face non-functional requirements. In particular, the authors were interested in collecting information about the identification and measurement mechanisms put in place, the importance of various non-functional requirements from their perspective, and the challenges associated with the identified non-functional requirements. This work can be therefore seen as complementary to our systematic literature review. On the one hand, we enlarge the knowledge of non-functional requirements in machine learning-enabled systems by synthesizing the current literature from various perspectives. On the other hand, we could corroborate and further extend the findings on the challenges identified through a different research method. 

Finally, Ahmad et al. \cite{ahmad2023requirements} recently performed a systematic mapping of the research on requirements engineering for artificial intelligence. The authors surveyed the approaches to specify requirements and the frameworks/tools/methods to model them. Our work is clearly complementary, as we focus on non-functional attributes, attempting to classify them and synthesize the current knowledge on the challenges they face.

While the papers discussed above are closely connected to the work proposed herein, it is also worth mentioning the existence of an ever-increasing number of secondary studies targeting multiple aspects of software engineering for artificial intelligence research. 

A consistent amount of systematic literature reviews focused on the synergies between the artificial intelligence and software engineering research communities \cite{wang2020synergy}, other than on the software engineering challenges and solutions for developing artificial intelligence systems \cite{kumeno2019sofware,lwakatare2020large,giray2021software,nascimento2020software}. With respect to them, our work focuses on non-functional requirements, hence providing finer-grained pieces of information and insights on the next steps that researchers should consider to better support practitioners. 

Other researchers targeted the quality assurance problem, which is deemed one of the most relevant and complex for the SE4AI research community. In particular, we identified systematic literature review and empirical investigations into the field of software quality \cite{masuda2018survey,gezici2022systematic}, design patterns \cite{washizaki2019studying,serban2020adoption},
software architecture \cite{serban2021empirical}, and testing \cite{borg2018safely,braiek2020testing,riccio2020testing,zhang2020machine}. Finally, recent systematic literature reviews have been conducted with the aim of understanding the deployment strategies for artificial intelligence systems \cite{john2021architecting}, and model-based development approaches \cite{lorenzoni2021machine}.

The substantial efforts put by researchers in conducting systematic investigations on software engineering for artificial intelligence further motivate our work, as they let arise the collective, emerging interest around these matters. Our systematic literature review aims, therefore, to provide additional insights to the research community by studying the non-functional requirements of machine learning-enabled systems that have been studied over the last years.

%% file: sections/3_Methodology.tex
\section{Research Method}
\label{sec:methodology}
The ultimate \emph{goal} of the study was to systematically survey the current research on non-functional requirements of machine learning-enabled systems, with the \emph{purpose} of providing researchers in the field of SE4AI with actionable items and insights that they can exploit to further explore the matter and improve the support provided to machine learning engineers. Through our work, we therefore provided a \emph{synthesis} of the state of the art rather than a \emph{comparison} between the state of the art and the state of the practice: such a comparison would be undoubtedly beneficial, yet is deemed as part of our future research agenda. Specifically, the research was conducted in terms of (1) the classification of the non-functional requirements of ML-enabled systems, and (2) the challenges that non-functional requirements pose during the development of ML-enabled systems. The \emph{perspective} is of fresh Ph.D. students, senior researchers, and practitioners. The former are interested in mapping the research literature that focuses on addressing non-functional requirements in machine learning-enabled systems so that they can possibly identify neglected research areas to further investigate. Senior researchers are interested in mapping the state of the art in order to identify opportunities for research proposals and industrial collaborations. Finally, practitioners are interested in understanding the challenges to face when optimizing non-functional requirements, which may possibly let arise the obstacles to technological transfer that should be carefully considered in practice.

To address our goal, we designed and executed a systematic literature review (SLR), which is a process through which the existing scientific papers on a subject of interest are rigorously identified, selected, and analyzed to address one or more research questions \cite{kitchenham}. To make our SLR as complete as possible, we first followed the well-established guidelines proposed by Kitchenham et al. \cite{kitchenham2009systematic}. However, as further elaborated in Section \ref{searchString}, we had to preserve the sustainability of the data collection and analysis procedures when designing the search string. Specifically, we defined a search string that looked for articles that explicitly referred to terms like \emph{``Non-Functional Requirements''}, possibly missing the articles that referred to the specific non-functional requirements considered, e.g., \emph{``fairness''}. To mitigate threats due to incompleteness of the search, we boosted the search process by means of two additional steps. On the one hand, we systematically screened the research articles published in top-tier software engineering and artificial intelligence venues in an effort to identify a set of seed papers to further process, i.e., the identification of seed papers has been shown to be an effective alternative when investigating research angles for which the standard guidelines do not enable a comprehensive search \cite{wohlin2020guidelines}. On the other hand, we provided additional rigor to the analysis by integrating the so-called snowballing procedure \cite{wohlin2016second}, i.e., the iterative scanning of the incoming and outcoming references of the primary studies done to identify additional relevant sources of information. Therefore, our systematic literature review can be considered \emph{``hybrid''} \cite{mourao2020performance}, namely a systematic analysis that combines traditional search strategies with additional search steps and snowballing. In terms of reporting, we followed the \textsl{ACM/SIGSOFT Empirical Standards}\footnote{The \textsl{ACM/SIGSOFT Empirical Standards}: \url{https://github.com/acmsigsoft/EmpiricalStandards}.} and, in particular, the \emph{``General Standard''} and \emph{``Systematic Reviews''} guidelines.


\subsection{Research Objectives and Questions}
The Goal-Question-Metric (GQM) approach \cite{caldiera1994goal} was initially employed to purposefully measure the goals to be achieved and relate them to the data analyzed. More specifically, the overall research objectives established by this systematic literature review were the following:

\steattentionbox{\noindent\faDotCircleO \hspace{0.05cm} \textbf{\underline{Objective 1.}} Providing a systematic classification of the non-functional requirements of ML-enabled software defined in the literature.
\\
\faDotCircleO \hspace{0.05cm} \textbf{\underline{Objective 2.}} Providing a systematic classification of the challenges to face when dealing with non-functional requirements of ML-enabled systems. 
}

Starting from these objectives, we defined more specific research questions (\textbf{RQ}s) that could drive our search process and analysis. These are presented in Table \ref{tab:RQs}. 

\begin{table}[ht]
    \centering
    \caption{Research Questions of the Systematic Literature Review.}
    \label{tab:RQs}
    \rowcolors{1}{gray!15}{white}
    \resizebox{\linewidth}{!}{
    \begin{tabular}{|p{2cm}|p{10cm}|} \hline 
    \rowcolor{black}
    \textcolor{white}{\textbf{Research Question}} & \textcolor{white}{\textbf{Description}} \\\hline
    \textbf{RQ$_1$} & What are the non-functional requirements of ML-enabled software considered by researchers?\\\hline
    \textbf{RQ$_2$} & What are the challenges of dealing with non-functional requirements of ML-enabled software?
    \\\hline
    \end{tabular}}
\end{table}

More particularly, \textbf{RQ$_1$} addressed the first objective of the study and aimed at systematically classifying the non-functional requirements that researchers have defined. By addressing this research angle, we aimed to contribute to previous research on the classification of non-functional requirements of ML-enabled systems \cite{habibullah2022non}. Indeed, while previous efforts in this respect have been conducted, there is still not a systematic classification - this represents a key contribution to our work. The second objective was mapped onto \textbf{RQ$_2$}. This research question elicited the major challenges that non-functional requirements pose during the development of ML-enabled systems.
\textbf{RQ$_2$} was able to enlarge the body of knowledge on the challenges to face when dealing with non-functional requirements, possibly informing researchers and practitioners on the aspects to further consider during the development of these systems. 

\subsection{Research Method to Conduct the Systematic Literature Search}
\label{searchString}
As a first step to address the goals of our study, we applied the guidelines by Kitchenham et al. \cite{kitchenham} to identify primary studies targeting non-functional requirements of ML-enabled systems. 

\subsubsection{Research Query definition}
\label{query_definition}
To develop an effective search string, we first extracted relevant terms from the research questions, identifying the keywords from the \textbf{RQs} \cite{kitchenham}. For all the terms, we then elaborated on the alternative spellings and synonyms. Afterward, we used boolean operators to assemble the search string, i.e., we used the \textsl{`OR'} operator for the concatenation of alternative spellings and synonyms, while the \textsl{`AND'} operator for the concatenation of relevant terms. More specifically, we elaborated on the following search string:
\smallskip
\stesearchstringbox{Search String.}{((``Machine Learning'') \textit{OR} (``Artificial Intelligence'') \textit{OR} (``Deep Learning'') \textit{OR} (``Reinforcement Learning'') \textit{OR} (``Deep Neural Network'') \textit{OR} (``ml'') \textit{OR} (``ai'') \textit{OR} (``dl'')) AND ((``nfr'') \textit{OR} (``Non-Functional Requirement*'') \textit{OR} (``Non Functional Requirement*''))}

{There are some considerations to make on the search string. In the first place, it included terms connected to machine, deep, and reinforcement learning, but also to deep neural networks and artificial intelligence. This was done to deal with the lack of a standard terminology: it is indeed possible that researchers used more generic terms, like \emph{``Artificial Intelligence''}, or more specific terms, like \emph{``Deep Neural Networks''} to indicate the analysis of ML-enabled systems.}

In the second place, the search string did not include terms related to any specific, known non-functional requirements, e.g., fairness, but focused on the more generic concept of non-functional requirement, including keywords like \emph{``Non-Functional Requirement''}, \emph{``Non Functional Requirement''}, and \emph{``NFR''}. We are aware that a search string including the specific non-functional requirements would have been ensured the collection of a larger amount of relevant papers. Nonetheless, we would have faced two issues. First, we would have been bound to the inclusion of the non-functional requirements explicitly classified in previous studies while defining the search string: as a consequence, we could not have satisfactorily addressed \textbf{RQ$_1$}, not being able to classify additional non-functional requirements emerging from primary studies. Perhaps more importantly, the systematic literature review would have become prohibitive in terms of effort. For instance, Habibullah et al. \cite{habibullah2021non} estimated the number of hits for a search query that includes all classified non-functional requirements in over 200,000 articles: on the one hand, such a search query would have had a low precision, identifying a high amount of irrelevant resources; on the other hand, the application of exclusion and inclusion criteria over such a large set of candidate articles would have required an excessive effort. As such, we had to identify an alternative solution,  opting for the implementation of a hybrid mechanism to search the additional relevant resources through the processes described in Sections \ref{seed_set} and \ref{snowballing}.

Our approach to the search string design is similar to one of the recent systematic literature reviews in the field of SE4AI proposed by Gezici and Tarhan \cite{gezici2022systematic}, and Martinez-Fernandez et al. \cite{martinez2022software}. 
Based on the arguments made above, we may conclude that we preferred to have a more generic search string in an effort to increase the recall of the search process---being able to collect a larger amount of papers---while preserving the sustainability of the overall data collection and analysis process. Of course, this decision impacted the precision of the search process and, subsequently, the effort required to apply the exclusion and inclusion criteria. Nonetheless, we accepted this compromise to make sure to include all relevant sources in our investigation.

\subsubsection{Search Databases}
\label{study_selection}
We applied the search on \textsl{ACM Digital Library},\footnote{Link to \textsl{ACM Digital Library}: \url{https://dl.acm.org}.} \textsl{Scopus},\footnote{Link to \textsl{Scopus}: \url{www.scopus.com}.} and \textsl{IEEEXplore}.\footnote{Link to \textsl{IEEEXplore}: \url{http://ieeexplore.ieee.org}.} These digital libraries are often used to carry out systematic literature and mapping studies, as they provide a comprehensive set of resources to conduct them. It is worth noticing that \textsl{Scopus} indexes all the papers published by relevant publishers such as \emph{Springer} and \emph{Elsevier} - this is the reason why we did not include the \textsl{SpringerLink}\footnote{Link to \textsl{SpringerLink}: \url{https://link.springer.com/}.} and \textsl{ScienceDirect}\footnote{Link to \textsl{ScienceDirect}: \url{https://www.sciencedirect.com}.} databases. At the same time, we still opted for the inclusion of \textsl{ACM Digital Library} and \textsl{IEEEXplore}. This was done on the basis of the fact that the proceedings of some relevant ACM and IEEE conferences might not have been indexed by \textsl{Scopus} and, for this reason, we might have missed relevant resources for our study. These databases only include peer-reviewed articles: as such, we did not need to double-check whether the papers extracted from them were peer-reviewed.


\subsubsection{Exclusion and Inclusion criteria}
\label{esclusion_inclusion}
The papers retrieved from the search process were assessed against the following exclusion and inclusion criteria \cite{kitchenham2009systematic}.

\begin{description}[leftmargin=0.3cm]
    \item[Exclusion criteria.] The resources that met the constraints reported below were excluded:

    \begin{itemize}
        \item Papers not written in English;
        \item Duplicated papers;
        \item Papers whose full-text read was not available;
        \item Paper not published or not peer-reviewed;
        \item Workshop, systematic, survey papers;
        \item Short papers, with a page count of less than five pages;
        \item Papers that did not fall into themes of computer science and computer engineering;
        \item Papers published before 2012;
        \item Conference papers which were later extended as journal submissions;
        \item Papers out of scope;
    \end{itemize}

    \smallskip
    \item[Inclusion criteria.] We included the resources that met at least one of the following constraints:

    \begin{itemize}
        \item Papers defining types of non-functional requirements of ML-enabled systems;
        \item Papers describing issues or challenges to face when dealing with non-functional requirements of ML-enabled systems;
        \item Papers presenting approaches or tools to analyze non-functional requirements of ML-enabled systems;
        \item Papers presenting methodologies to optimize non-functional requirements problems in ML-enabled systems.
    \end{itemize}

\end{description}

\subsection{Research Method to Conduct the Seed Set Search}
\label{seed_set}
As further detailed in Section \ref{sec:SLRexecution}, the research method implemented through the guidelines by Kitchenham et al. \cite{kitchenham} was found to be insufficient in finding relevant articles. Similarly to Habibullah et al. \cite{habibullah2021non}, we found out that the relevant papers might have referred to specific non-functional requirements, e.g., \emph{``fairness''}, rather than to the more generic term \emph{``non-functional requirement''}. As such, the traditional systematic literature review approach might not have ensured completeness. 


\begin{table}[ht]
    \caption{\revised{Conferences and journals considered in the scope of the seed set identification process.}}
    \label{table:list_conferences}
    
    \rowcolors{1}{gray!15}{white}
    \resizebox{\linewidth}{!}{
    \begin{tabular}{|c|c|p{13cm}|c|} \hline 
    \rowcolor{black}
    \textcolor{white}{\textbf{Venue}} &
    \textcolor{white}{\textbf{Type}} & \textcolor{white}{\textbf{Name}} & \textcolor{white}{\textbf{Ranking}} \\\hline
    \textit{Journal} & SE-related & \textit{IEEE Transactions on Software Engineering \textbf{(TSE)}.} &  \textbf{Q1}\\\hline
    \textit{Journal} & SE-related & \textit{ACM Transactions on Software Engineering and Methodology \textbf{(TOSEM)}.} &  \textbf{Q1}\\\hline    
    \textit{Journal} & SE-related & \textit{Empirical Software Engineering \textbf{(EMSE)}.} &  \textbf{Q1}\\\hline
    \textit{Journal} & SE-related & \textit{Journal of Systems and Software \textbf{(JSS)}.} &  \textbf{Q1}\\\hline
    \textit{Journal} & SE-related & \textit{Information and Software Technology \textbf{(IST)}.} & \textbf{Q1}\\\hline
    \textit{\revised{Journal}} & \revised{AI-related} & \textit{\revised{Journal of Engineering Applications of Artificial Intelligence \textbf{(EAAI)}.}} & \textbf{\revised{Q1}}\\\hline
    \textit{\revised{Journal}} & \revised{AI-related} & \textit{\revised{IEEE Transactions on Knowledge and Data Engineering \textbf{(TKDE)}}.} & \textbf{\revised{Q1}}\\\hline
    \textit{\revised{Journal}} & \revised{AI-related} & \textit{\revised{IEEE Transactions on Pattern Analysis and Machine Intelligence \textbf{(TPAMI)}}.} & \textbf{\revised{Q1}}\\\hline
    \textit{\revised{Journal}} & \revised{AI-related} & \textit{\revised{IEEE Transactions on Neural Networks and Learning Systems \textbf{(TNNLS)}}.} & \textbf{\revised{Q1}}\\\hline\hline
    
    \textit{Conference} & SE-related & \textit{IEEE/ACM International Conference on Software Engineering \textbf{(ICSE)}.} &  \textbf{A*}\\\hline
    \textit{Conference} & SE-related & \textit{Joint European Software Engineering Conference and the ACM SIGSOFT Symposium on the Foundations of Software Engineering \textbf{(ESEC/FSE)}.} &  \textbf{A*}\\\hline
    \textit{Conference} & SE-related & \textit{IEEE/ACM Automated Software Engineering Conference \textbf{(ASE)}.} &  \textbf{A*}\\\hline
    \textit{Conference} & SE-related & \textit{IEEE International Conference on Software Maintenance and Evolution \textbf{(ICSME)}.} & \textbf{A}\\\hline
    \textit{Conference} & SE-related & \textit{IEEE International Conference on Software Testing, Verification and Validation \textbf{(ICST)}.} & \textbf{A}\\\hline
    \textit{Conference} & SE-related & \textit{ACM SIGSOFT International Symposium on Software Testing and Analysis \textbf{(ISSTA)}.} & \textbf{A}\\\hline
    \textit{Conference} & SE-related & \textit{IEEE International Requirements Engineering Conference \textbf{(RE)}.} & \textbf{A}\\\hline
    \textit{Conference} & SE-related & \textit{IEEE International Conference on Software Analysis, Evolution and Reengineering \textbf{(SANER)}.} & \textbf{A}\\\hline
    \textit{Conference} & SE-related & \textit{IEEE International Conference on Program Comprehension \textbf{(ICPC)}.} & \textbf{A}\\\hline
    \textit{Conference} & SE-related & \textit{IEEE International Working Conference on Mining Software Repositories \textbf{(MSR)}.} & \textbf{A}\\\hline
    \textit{\revised{Conference}} & \revised{AI-related} & \textit{\revised{The Association for the Advancement of Artificial Intelligence \textbf{(AAAI)}.}} & \textbf{\revised{A*}}\\\hline
    \textit{\revised{Conference}} & \revised{AI-related} & \textit{\revised{International Joint Conference on Artificial Intelligence \textbf{(ICAI)}.}} & \textbf{\revised{A*}}\\\hline
    \textit{\revised{Conference}} & \revised{AI-related} & \textit{\revised{Empirical Methods in Natural Language Processing \textbf{(EMNLP)}.}} & \textbf{\revised{A*}}\\\hline
    \end{tabular}}
\end{table}

To address this problem, we conducted the so-called \emph{seed set identification}. This is the process through which a researcher systematically screens the research papers published in top conferences and journals over a given time period in an effort to identify additional resources relevant to the research they are conducting. Wohlin et al. \cite{wohlin2020guidelines} recently showed that this approach could be used to effectively complement a systematic literature review in cases where the standard guidelines cannot lead to a comprehensive search. In our context, the seed search process should have considered both software engineering and artificial intelligence venues, as studies targeting the analysis of non-functional requirements and/or the design of automated approaches to deal with them might have been published in both the research areas.

In the case of software engineering venues, we identified the most relevant software engineering conferences and journals, relying on (1) the \textsl{CORE} Ranking system\footnote{The \textsl{Computing Research and Education Association of Australasia}: \url{https://www.core.edu.au/}} to select A* and A conferences, and (2) the \textsl{Scimago} Journal Ranking\footnote{The \textsl{Scimago} Journal Ranking: \url{https://www.scimagojr.com}.} to identify the software engineering journals falling into the first quartile of all journals (Q1) in the \textsl{`Software'} category. As such, we selected the conferences and journals marked as \textsl{`SE-related'} presented in Table \ref{table:list_conferences}. In the second place, we scanned all the papers published at those conferences starting from 2022 backward until 2012. 
As a final step, we analyzed each conference website, and for journals, we searched for papers via the \textsc{DBLP},\footnote{The \textsl{DBLP}: \url{https://dblp.org/}.} the most extensive computer science bibliography library. We applied the same exclusion and inclusion criteria defined for the search process conducted using the standard guidelines (see Section \ref{esclusion_inclusion}).

In the case of artificial intelligence venues, opening up to all the most relevant ones would have been prohibitively expensive in terms of effort. Using the same criteria as for the software engineering venues, we would have selected 70 Q1 journals and more than 40 A* and A conferences, which we would have scanned from 2022 backward until 2012. We estimated the potential hits to analyze in around 200,000, which was deemed overly expensive for a human assessment. Nonetheless, to make our systematic literature process as complete as possible, we narrowed down the scope of the seed search to the artificial intelligence venues having a higher likelihood to publish engineering or empirical approaches to the development of ML-enabled systems, i.e., the venues that are more likely to contain papers relevant to our research questions.

In particular, we first identified an online repository, named \textsl{`AI Venues'},\footnote{The \textsl{`AI Venues'} repository: \url{https://aivenues.github.io/}}, which lists the whole set of artificial intelligence conferences and journals along with their ranking and H-index. We then associated to each journal the corresponding rank provided by \textsl{Scimago} and to each conference the rank provided by \textsl{CORE}. Besides discarding the venues that did not meet our selection criteria (rank=A* or A for conferences, rank=Q1 for journals), we decided to filter out the venues that revolved around too specific techniques or technologies, e.g., the \emph{IEEE Transactions on Image Processing} journal, and favor instead the venues that encompassed a broader spectrum of engineering or empirical approaches applied for the development of ML-enabled systems, e.g., the \emph{IEEE Transactions on Neural Networks and Learning Systems}. In this way, we could have focused on papers that had higher likelihood to be relevant for our research questions. Based on this process, we selected four Q1 journals and three A* conferences whose themes were either related to the improvement of AI approaches or the use of engineering or empirical approaches to AI. This process led to the selection of the venues marked as \textsl{`AI-related'} reported in Table \ref{table:list_conferences}. Also in this case, we scanned all the papers published to these venues between 2012 and 2022, applying the same selection process described in Section \ref{esclusion_inclusion}.

The seed search identified additional primary studies published to the selected venues. As a consequence, we did not need to double-check whether these were actually published or not.

\subsection{Research Method to Conduct the Snowballing Process}
\label{snowballing}
The third step conducted revolved around the so-called forward and backward snowballing. This is the procedure through which a researcher systematically scans the incoming and outcoming references of the primary studies with the aim of identifying new relevant resources to address the research questions \cite{10.1145/2601248.2601268}. In our case, the primary studies identified as a consequence of the application of the exclusion and inclusion criteria on both the studies retrieved using the standard guidelines and the seed search were scanned. In particular, we applied an iterative procedure in which all incoming and outcoming references of the primary studies were first considered. Afterward, we reiterated the procedure for the newly acquired studies in an effort to identify additional resources. Overall, four rounds of backward and four rounds of forward were conducted: we stopped at four as we reached saturation, namely, we did not identify any additional papers to include. From a practical perspective, the snowballing steps were conducted through the use of \textsl{Google Scholar},\footnote{Link to \textsl{Google Scholar}: \url{https://scholar.google.com/}} an academic search engine which simplifies the analysis of incoming and outcoming references of research papers. This was the only step where we actually needed to verify whether the papers cited or citing the primary resources were published. Whenever needed, i.e., whenever we identified a new paper which was not previously identified through the initial or the seed search, we searched the title of the paper on \textsl{ACM Digital Library}, \textsl{Scopus}, and \textsl{IEEEXplore} to verify its publishing status. In the case the paper was published, it was accepted for the subsequent steps of our research method.

\subsection{Research Method to Conduct the Quality Assessment}
Once we had completed the application of the three complementary search processes described in the previous sections, we conducted a quality assessment of the resources that successfully passed the inclusion criteria. 
The implementation of the quality assessment process started with the definition of qualitative questions aiming at operationalizing the main pieces of information that a primary study should have had in order to be useful to address our research questions. These were defined according to the research objectives initially defined:

\begin{itemize}
    \item \textbf{Q1:} \emph{Is the non-functional requirement well defined?}
    \item \textbf{Q2:} \emph{If the primary study defined a challenge to deal with a non-functional requirement, is this well defined?}
    \item \textbf{Q3:} \emph{If the primary study proposes an automated approach to deal with a non-functional requirement, was the engineering challenge(s) faced behind the definition of the approach clearly defined?}
    \item \textbf{Q4:} \emph{If the primary study proposes a methodology to optimize non-functional requirements, was the engineering challenge(s) faced behind the definition of the methodology clearly defined?}
    
\end{itemize}

In particular, \textbf{Q1} was the instrumental to ensure a proper answer to \textbf{RQ$_1$}, while \textbf{Q2}, \textbf{Q3}, and \textbf{Q4} were used to address \textbf{RQ$_2$}. As shown, we did not limit ourselves to the papers that explicitly analyzed the challenges to face when dealing with non-functional requirements, but also elicited those challenges by means of a qualitative assessment of the motivations behind the introduction of automated approaches and optimization methodologies. 

When assessing the primary studies against each of the qualitative questions, previous systematic literature reviews (e.g., \cite{ahmad2023requirements,azeem2019machine,de2018systematic}) assigned a boolean value to indicate whether a study had or not the quality required with respect to a property considered. However, assessing the primary studies through boolean values might be challenging, other than possibly threatening the validity of the assessment. For instance, there might be cases where the challenges associated with non-functional requirements may be logically elicited from the text, even though not explicitly stated. To deal with this process, we, therefore, opted for the application of a \emph{fuzzy linguistic approach} \cite{https://doi.org/10.1002/smr.2211}, which consists of rating each primary study through a continuous variable ranging between 0 and 1. In particular, the scores were assigned as follows:

\begin{multicols}{5}
\small
\begin{itemize}
    \item 0 $\Rightarrow$ \textbf{No}
    \item 0.1-0.3 $\Rightarrow$ \textbf{Rarely}
    \item 0.4-0.6 $\Rightarrow$ \textbf{Partly}
    \item 0.7-0.9 $\Rightarrow$ \textbf{Mostly}
    \item 1 $\Rightarrow$ \textbf{Yes} 
\end{itemize}
\end{multicols}

In other terms, for each qualitative question, the primary studies were assigned a value reporting how explicit and clear the content was in that respect. At the end of the evaluation conducted for each question, the total merit of a primary study was computed as follows:

\begin{multicols}{5}
\small
\begin{itemize}
    \item 0 $\Rightarrow$ \textbf{No}
    \item 0.1-1.5 $\Rightarrow$ \textbf{Rarely}
    \item 1.6-2.8 $\Rightarrow$ \textbf{Partly}
    \item 2.9-3.6 $\Rightarrow$ \textbf{Mostly}
    \item 3.7+ $\Rightarrow$ \textbf{Yes} 
\end{itemize}
\end{multicols}

To be finally accepted as part of our systematic analysis, the primary study should have obtained a final score of more than 1.6, i.e., it should have partly specified the required pieces of information to address the research questions of the study. 
 
\subsection{Design of the Data Extraction Form}
\label{data_extraction}
As a final step of the research method applied to address the goals of our study, we designed the data extraction form, namely the specification of the pieces of information to collect when addressing our research questions. Table \ref{table:data_extraction} summarizes the data collected, reporting (i) the dimension the attribute group referred to; (ii) the scope where the data has been used; (iii) the description of the dimension considered; and (iv) the specific attributes considered. As shown in the table, we collected six main categories of information. At first, we identified pieces of information that might help statistically describe our sample in terms of bibliometrics, e.g., publication year and venues: we used the knowledge acquired to describe the trends in terms of publication, the most relevant venues accepting research papers on non-functional requirements of ML-enabled systems. Besides these meta-data, we then collected and stored information that may be directly connected to the specific research questions posed in our study and that, therefore, was used in the context of the data analysis and reporting process. 


\begin{table}[ht]
    \caption{Data extraction Form.}
    \label{table:data_extraction}
    
    \rowcolors{1}{gray!15}{white}
    \resizebox{\linewidth}{!}{
    \begin{tabular}{|p{6cm}|l|p{10cm}|p{6cm}|} \hline 
    \rowcolor{black}
    \textcolor{white}{\textbf{Dimension}} & \textcolor{white}{\textbf{Scope}} & \textcolor{white}{\textbf{Description}} & \textcolor{white}{\textbf{Attribute Collected}}\\\hline
    Paper Information & \textbf{Bibliometrics} & This component includes general information and criteria used to assess the quality of the selected studies. & Seed set or Systematic Literature Review\\
    \cline{4-4}
        & & & Year of publication \\   
    \cline{4-4}
        & & & Accepted Score \\
    \cline{4-4}
        & & & Journal or a conference. \\\hline
    Terms concerning non-functional requirements & \textbf{RQ$_1$} & This component encompasses all the keywords that make it possible to describe non-functional requirements, both those that already exist in the literature and new non-functional requirements that emerged from this study. & Non-Functional Requirements \\\hline 
    Domains where non-functional requirements were studied & \textbf{RQ$_1$} & This component is similar to row 2 but contains all the domains in which non-functional requirements have been studied and found to be problematic. This will make it possible to diversify the impact of AI according to context and to identify the least researched areas. & ML Domains\\
    \cline{4-4}
    & & & Environmental Domains \\\hline
    Challenges of SE approaches for ML-enabled systems & \textbf{RQ$_2$} & This component contains a list of challenges explicitly stated in primary studies and any future challenges and problems emerging from these studies. This information will help us identify areas where further research is needed to improve SE approaches for ML-enabled systems and improve the existing ones. & Challenges \\
    \cline{4-4}
        & & & New Challenges Emerged \\\hline
    \end{tabular}}
\end{table}


\begin{figure}[ht]
\includegraphics[width=1\linewidth]{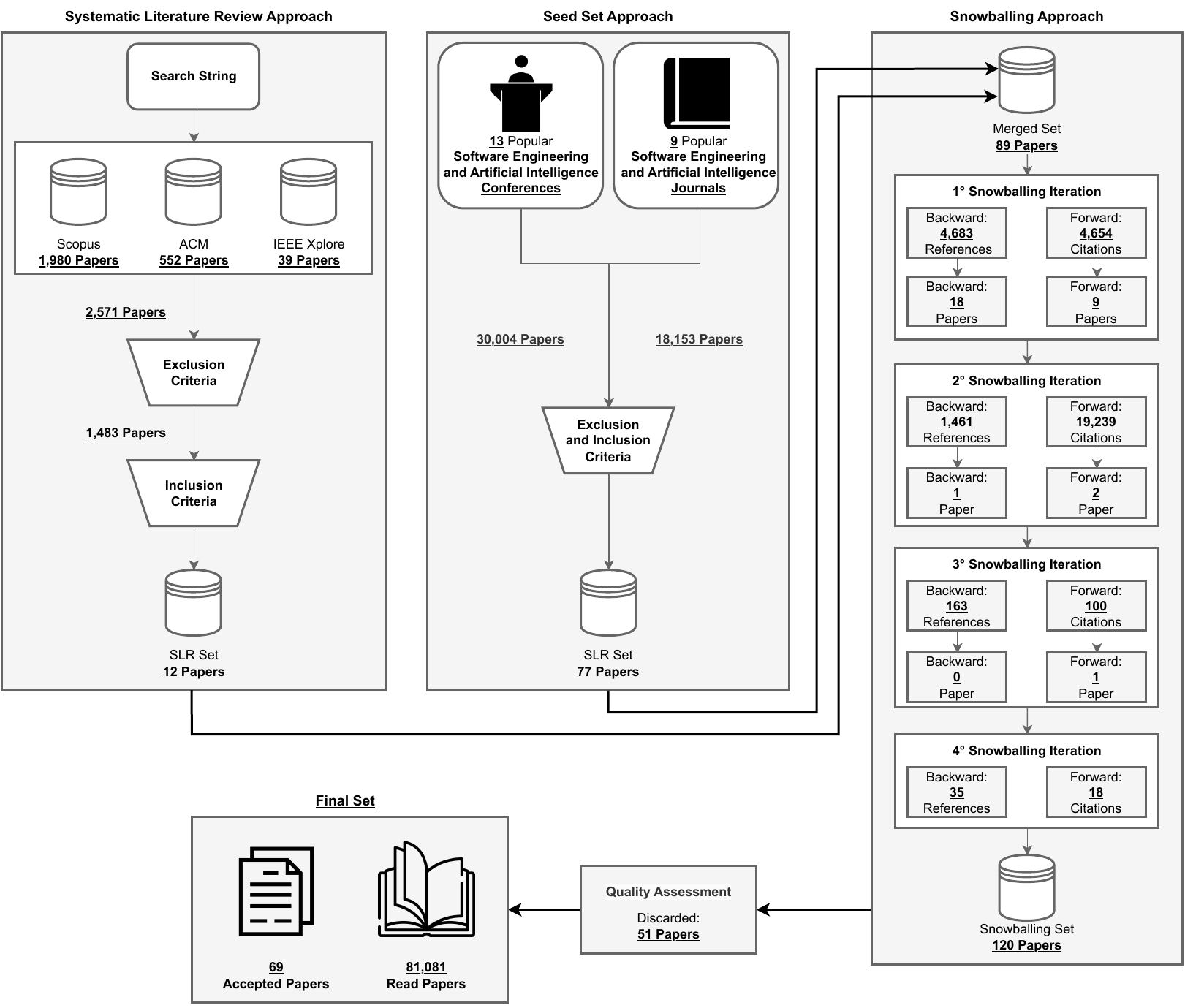}
\caption{Overview of the process for selecting papers.}
\label{fig:fig1}
\end{figure}

\subsection{Execution of the Research Methods}
\label{sec:SLRexecution}
Figure \ref{fig:fig1} overviews the outputs of the execution of the research methods designed in the previous sections. The systematic literature review search was first conducted by the first author of the paper on September 4, 2023. As summarized in the figure, the application of the search query produced a total amount of 2,571 hits. As expected, most of these hits came from \textsl{Scopus} (1,980), with \textsl{ACM Digital Library} and \textsl{IEEEXplore} reporting a lower amount of results (552 and 39, respectively). 

The first author of the paper manually collected all the relevant information on the resources identified and stored them within a sheet that was later shared with the second author. Afterwards, the first author proceeded with the application of exclusion and inclusion criteria: the former led to filtering out 1,088 research results, while the latter had a total of 1,471. The very limited amount of relevant papers identified after the application of the guidelines by Kitchenham et al. \cite{kitchenham} led the two authors to open a discussion about the method defined and the implications that such an outcome might have had on the completeness of the search. In particular, the two authors scheduled two meetings of one hour each where they jointly reviewed the activities performed in an effort to understand the motivations behind the limited amount of resources retrieved. As a result of this fine-grained investigation, they could draw two main observations:

\begin{enumerate}
    \item The vast majority of the hits were found to be irrelevant to the goals of the study because they mentioned terms connected to non-functional requirements or machine learning either as part of the background or the related work, hence not discussing the challenges posed by non-functional requirements in the context of ML-enabled systems;

    \smallskip
    \item The few relevant hits which passed the inclusion criteria did not only explicitly mention the terms of our search string but also discussed the fine-grained requirements considered. 
    
\end{enumerate}

Based on these observations, the two authors realized the need for boosting the traditional search process with additional steps. While they first considered the re-execution of the systematic search through a search string designed to include all non-functional requirements classified by researchers so far, this would have led to three critical issues. First, the answer to \textbf{RQ$_1$} would have been biased by design: we would have indeed introduced ourselves to the set of non-functional requirements to look for rather than letting them emerge from the analysis of the available literature. Second, the search string would have included a number of \textsl{AND}/\textsl{OR} conditions that might have increased noise, leading to the retrieval of several resources not connected to SE4AI research. Last but not least, such an increased noise would have led to the identification of a prohibitively expensive amount of resources to analyze---Habibullah et al. \cite{habibullah2021non} estimated over 200,000 hits for a search string including all the non-functional requirements of ML-enabled systems classified in literature. 

To cope with these challenges, we opted for complementing the systematic literature search with the seed set identification and the snowballing process. Since our goal was that to analyze the literature available in the field of SE4AI on non-functional requirements, we then proceeded with the analysis of the papers published to top-tier software engineering and artificial intelligence venues. The first author of the paper looked for those resources, extracting information concerned to 30,004 papers from conferences and 18,153 papers from journals. 
After applying the exclusion and inclusion criteria, 77 unique new resources were deemed relevant, 71 from conferences and 6 from journals.

Before proceeding with the next steps, the second author of the paper reviewed the activities conducted so far. Also, two one-hour meetings were scheduled to discuss the results achieved. The meetings allowed us to acknowledge the soundness of the decision made when opting for a seed search. Hence, the 89 relevant resources acquired so far were taken as input for the snowballing step. Also, in this case, the first author was responsible for conducting the search process: by iterating multiple times, both backward and forward, the author was able to identify an additional set of 31 primary studies. Finally, the total amount of 120 papers identified through the three-step search process was considered within the scope of the quality assessment. Given the criticality of this step, this was jointly conducted by the authors: after analyzing 81,081 results, 69 primary studies finally contributed to our study.


To make the reader aware of the effort required to conduct the search/analysis process and contribute to the transparency/replicability of our work, we also estimated the number of person-hours invested in the work. As for the first author, the estimation is about 800 person-hours; as for the second, it is about 400 person-hours.

\smallskip
\stesearchstringbox{Reproducibility and Replicability of Our Work.}
{Our online appendix reports all the fine-grained steps described above. The interested reader may use the appendix to replicate our work or build on top of our findings to create additional knowledge on non-functional requirements of ML-enabled systems: \cite{DeMartino2023}}

%% file: sections/4_Results.tex
\section{Analysis of the Results}
\label{sec:results}
In this section, we report quantitative and qualitative insights coming from the data extraction and analysis phase. We first analyze bibliometric data to describe the sample considered in our research. Afterward, we address the specific research questions of the study.

\begin{figure}[htbp]
  \centering
  \begin{minipage}[b]{0.45\textwidth}
    \includegraphics[width=\textwidth]{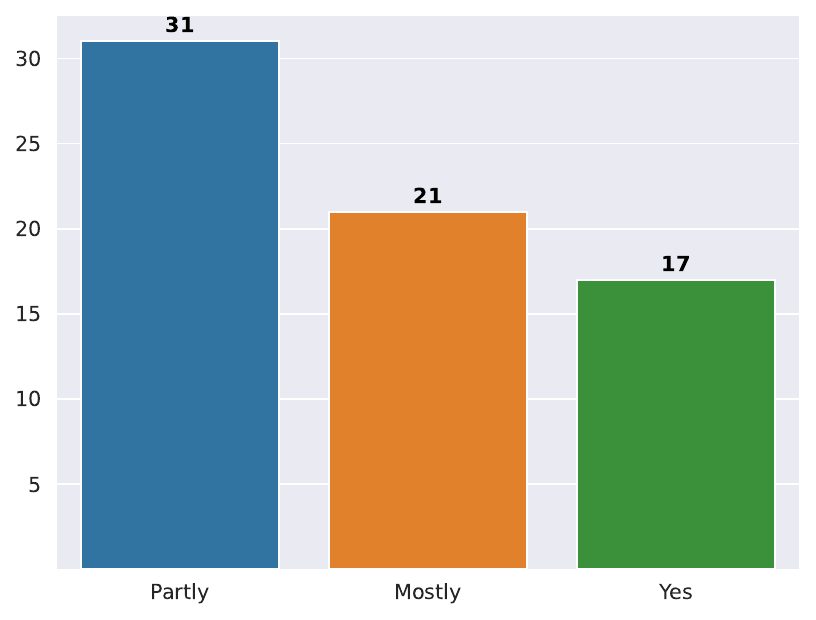}
    \caption{Results of the quality assessment.}
    \label{fig:n_score}
  \end{minipage}
  \begin{minipage}[b]{0.45\textwidth}
    \includegraphics[width=\textwidth]{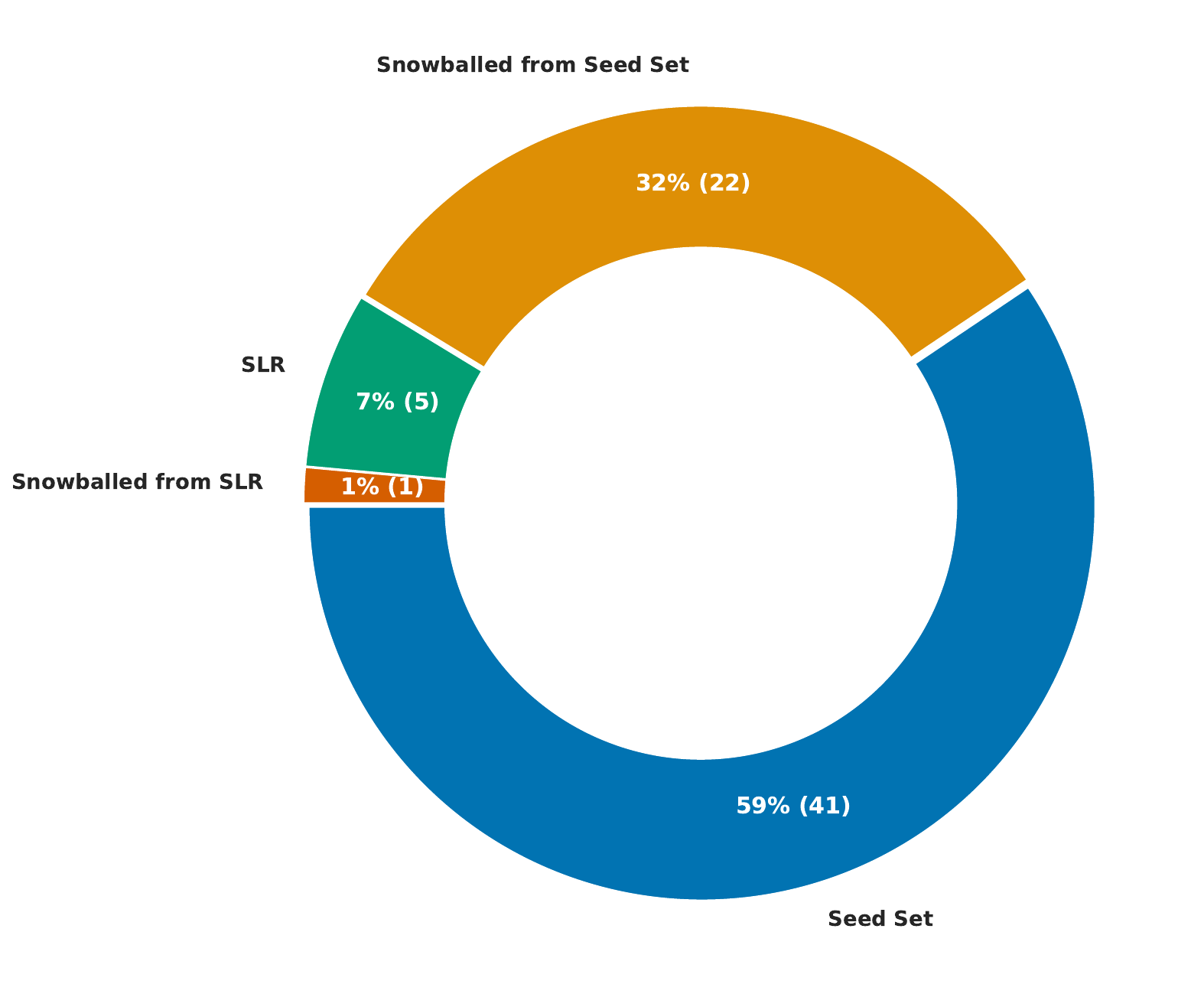}
    \caption{Papers per publication venue.}
    \label{fig:sources}
  \end{minipage}
\end{figure}

\subsection{Bibliometrics}
Figure \ref{fig:n_score} overviews the results of the quality assessment procedure. We finally included a total of 69 primary studies. Most of them (31, 45\%) received an overall quality score of \textsl{``Partly''}, i.e., they specified the pieces of information required to address our research questions mostly implicitly, yet giving us a chance to elicit them satisfactorily. The other 38 primary studies were more explicit and, indeed, reached an overall quality score of \textsl{``Mostly''} or \textsl{``Yes''}. The primary studies were mostly retrieved through the seed set search (41, 59\%), while the other 22 resources (32\%) were identified by snowballing the seed primary studies identified. Only 5 papers were retrieved through the systematic literature search and only one additional resource could be identified through the snowballing process conducted on the set of papers identified with the systematic search. These considerations further justified our choice of complementing the traditional systematic literature search with additional instruments. 

\begin{figure}[htbp]
  \centering
  \begin{minipage}[b]{0.45\textwidth}
    \includegraphics[width=\textwidth]{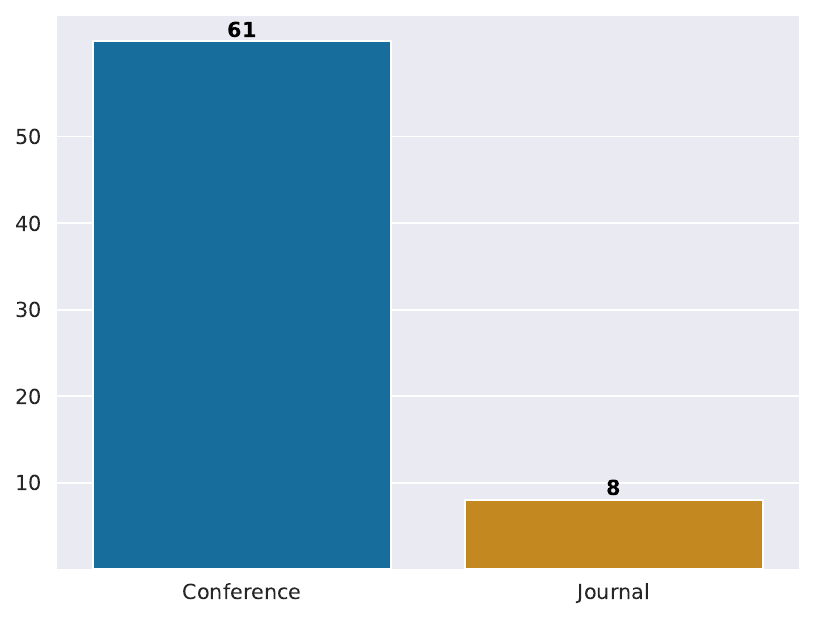}
    \caption{Papers published per publication venue.}
    \label{fig:conference_journal}
  \end{minipage}
  \begin{minipage}[b]{0.45\textwidth}
    \includegraphics[width=\textwidth]{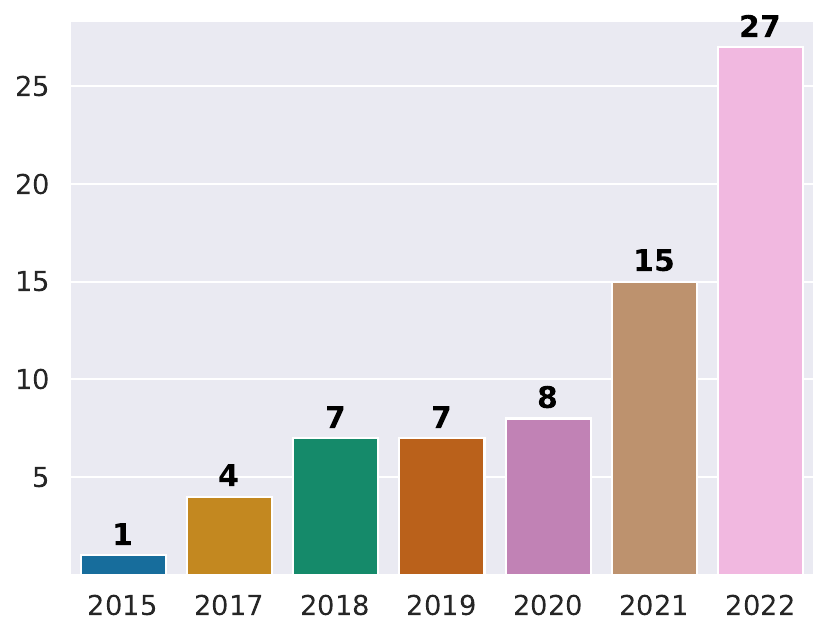}
    \caption{Papers published on the matter over time.}
    \label{fig:year_paper}
  \end{minipage}
\end{figure}

More interesting were the insights coming from the publication venues, which are shown in Figure \ref{fig:conference_journal}. We observed that the vast majority of the primary studies (61, 88\%) appeared in conference proceedings, with a limited amount of resources published in journals (8, 12\%). This finding seems to suggest that the research on non-functional requirements of ML-enabled systems is still at its early stage, with researchers interested in discussing recent advances in venues that allow discussion and interaction with the research community. On a similar note, the result may stimulate young researchers to enter this growing research area. The analysis of the amount of papers published on the matter per year confirmed the early nature of the research area (Figure \ref{fig:year_paper}). According to our findings, until 2015, no articles were published, and from 2015 to 2018 only a few papers were published, while a notably increasing trend could be found in the last two years. Performing a systematic synthesis of the knowledge collected so far could provide a relevant boost to the research activities that will be performed in the next years: our work indeed identifies multiple implications and challenges that the software engineering research community will be called to address. 



\subsection{\textbf{RQ$_1$} - What are the non-functional requirements of ML-enabled software considered by researchers?} 
The first goal of our work was to classify the non-functional requirements of machine learning-enabled systems. The primary studies considered were first analyzed in an effort to elicit the non-functional attributes that they aimed to address. Afterward, we provided the extracted non-functional attributes with a (1) name, (2) a description, and (3) a reference class, i.e., a high-level class that may be used to cluster together multiple non-functional requirements based on similarity, i.e., non-functional requirements sharing similar purposes. For instance, requirements such as \emph{``privacy''} and \emph{``security''} were both extracted and mapped onto the \textsl{``Resiliency''} class, as they both referred to aspects making an ML-enabled system resilient to unexpected events arising in operation, including external attacks. Whenever possible, i.e., when considering the papers that received a quality score of \textsl{``Yes''}, the extraction was relatively straightforward as these resources explicitly referred to the non-functional requirements considered; as such, they could be easily extracted and mapped onto known classes of non-functional requirements or onto new classes according to the conceptual similarity between the non-functional requirement analyzed and the set of classes identified. In the other cases, relevant terms connected to non-functional aspects were extracted, elaborated, and only later mapped onto the corresponding class of non-functional requirements. For example, the primary study \citeS{10.1145/3530019.3530025} reported terms like \emph{``memory problems''} and \emph{``battery drain''}, which we used to interpret the context of the study and assigned the non-functional aspect to the class entitled \textsl{``Sustainability''}. As an output of the process, we were able to develop a taxonomy of non-functional attributes of ML-enabled systems. 

It is worth pointing out that the names assigned to the classes, the description of each non-functional requirement, and the classification exercise as a whole, were informed by different sources of the existing body of knowledge, i.e., by the primary studies collected in this paper, other resources of the state of the art not directly related to our research goals, and the \textsl{Systems \& Software Quality Requirements \& Evaluation} standard (ISO/EIC 25010). In particular, the process was initially conducted by the first author of the paper, who (1) analyzed the primary studies to elicit the non-functional attributes considered and (2) provided a preliminary version of the taxonomy. The author exploited the existing body of knowledge in an effort to homogenize the definitions and classification; e.g., he favored the assignment of well-established class names whenever possible in an effort to provide a more comprehensive and usable taxonomy. The preliminary taxonomy was later the subject of evaluation. The second author joined the process at this stage, and a discussion was opened on the taxonomy produced. The two authors discussed about the consistency of the classification performed, other than on the names and descriptions assigned. Whenever needed, they modified the taxonomy---this happened in four cases, in which some non-functional requirements were grouped differently, and the names assigned to the classes were modified. 


In the first place, as a result of this procedure we could identify a total amount of 30 non-functional requirements. These were: \emph{accuracy} \citeS{usman2021nn,9439863,9508369,9623049,Eykholt_2018_CVPR,10.1145/3132747.3132785,10.1145/2810103.2813687,10.1145/3460120.3484818,8746775,8952197,10.1145/3551349.3556920,10.1145/3540250.3549093,10.1145/3338906.3338954,10.1145/3540250.3549103,10.1145/3510003.3510163,10.1145/3510003.3510087,10.1145/3510003.3510221,9402020,10.1145/3510003.3510231,10.1145/3510003.3510202,8812047,9402039,10.1145/3377811.3380368,10.1145/3510003.3510051,9402057,10.1145/3533767.3534373,10.1145/3533767.3534391,10.1145/3533767.3534408,10.1145/3533767.3534386,10.1145/3460319.3464809,secrypt19,9590404,10.1145/3368089.3409697,10.1145/3468264.3468565,9676691,10.1145/3506695,10.1145/3429444,10.1145/3243734.3243792,10.1145/3540250.3549102,10.1145/3540250.3549169,10.1145/3468264.3468537,10.1145/3368089.3409730,10.1145/3368089.3409668,10.1145/3368089.3409739,10.1145/3368089.3417065,10.1145/3180155.3180220,10.1145/3238147.3238187,10.1145/3510003.3510191,10.1145/3551349.3556906,10.1145/3510003.3510088,10.1145/3551349.3561158,Chen_2022_CVPR,10.1145/3340531.3411980}, \emph{robustness} \citeS{7958570,8418593,usman2021nn,9439863,9508369,9623049,Eykholt_2018_CVPR,10.1145/3132747.3132785,huang2017safety,9402124,9451178,9766323,10.1145/3460120.3484818,8746775,10.1145/3551349.3556920,10.1145/3338906.3338937,9402020,10.1145/3510003.3510231,8812047,9402039,10.1145/3377811.3380368,10.1145/3293882.3330580,10.1145/3533767.3534373,10.1145/3533767.3534391,10.1145/3533767.3534408,10.1145/3533767.3534386,10.1145/3460319.3464809,10.1145/3540250.3549102,10.1145/3368089.3409739,10.1145/3368089.3417065,10.1145/3238147.3238165,10.1145/3377811.3380331,10.1145/3238147.3238187,10.1145/3510003.3510191,10.1145/3510003.3510088,Chen_2022_CVPR}, \emph{security} \citeS{7958570,8418593,usman2021nn,9439863,10.1145/3132747.3132785,9402124,9451178,9766323,10.1145/3460120.3484818,10.1145/3551349.3556920,10.1145/3338906.3338954,9402020,8812047,9402039,10.1145/3377811.3380368,10.1145/3293882.3330580,10.1145/3533767.3534373,10.1145/3460319.3464809,secrypt19,9590404,9676691,10.1145/3368089.3409739,10.1145/3510003.3510191}, \emph{performance} \citeS{10.1145/3530019.3530025,9402124,9451178,9766323,8746775,10.1145/3468264.3468536,10.1145/3540250.3549093,10.1145/3540250.3549103,10.1145/3510003.3510163,10.1145/3510003.3510087,10.1145/3533767.3534391,10.1145/3533767.3534408,10.1145/3533767.3534386,secrypt19,10.1145/3506695,10.1145/3243734.3243792,10.1145/3540250.3549102,10.1145/3368089.3417065,10.1145/3551349.3556906,10.1145/3551349.3561158,Chen_2022_CVPR}, \emph{fairness} \citeS{7961993,pmlr-v81-buolamwini18a,10.1145/3338906.3338937,10.1145/3468264.3468536,10.1145/3540250.3549093,10.1145/3540250.3549103,10.1145/3510003.3510087,10.1145/3510003.3510202,9402057,9590404,10.1145/3368089.3409697,10.1145/3468264.3468565,10.1145/3540250.3549169,10.1145/3468264.3468537,10.1145/3238147.3238165,10.1145/3377811.3380331,10.1145/3340531.3411980,10.1145/3447548.3467177,10.1145/3534678.3539145}, \emph{behavioral} \citeS{10.1145/3132747.3132785,9451178,10.1145/3533767.3534391,10.1145/3338906.3338937,10.1145/3338906.3338954,10.1145/3510003.3510231,10.1145/3368089.3409730,10.1145/3180155.3180220,10.1145/3238147.3238165,10.1145/3238147.3238187,10.1145/3510003.3510191}, 
\emph{interpretability} \citeS{9451178,10.1145/3338906.3338954,10.1145/3510003.3510087,10.1145/3533767.3534373,9590404,10.1145/3429444,10.1145/3243734.3243792,10.1145/3368089.3409739},
\emph{transferability} \citeS{7958570,huang2017safety,9402124,9766323,10.1145/3377811.3380368,10.1145/3460319.3464809,10.1145/3510003.3510191}, \emph{safety} \citeS{Eykholt_2018_CVPR,10.1145/3132747.3132785,huang2017safety,10.1145/3338906.3338954,10.1145/3377811.3380368,10.1145/3180155.3180220,10.1145/3551349.3556906}, \emph{reliability} \citeS{9451178,10.1145/3510003.3510231,10.1145/3293882.3330580,10.1145/3533767.3534373,10.1145/3533767.3534391,9590404,10.1145/3238147.3238187}, \emph{explainability} \citeS{10.1145/3460120.3484818,10.1145/3338906.3338937,10.1145/3510003.3510202,10.1145/3429444,10.1145/3243734.3243792,10.1145/3447548.3467177},  \emph{retrainability} \citeS{9439863,9623049,9402039,10.1145/3293882.3330580,10.1145/3368089.3417065,10.1145/3377811.3380331}, \emph{cost} \citeS{10.1145/3510003.3510221,8812047,9590404,10.1145/3506695,10.1145/3368089.3417065}, \emph{energy consumption} \citeS{10.1145/3530019.3530025,10.1145/3510003.3510221,10.1145/3506695,10.1145/3540250.3549102,10.1145/3510003.3510088,10.1145/3551349.3561158}, \emph{privacy} \citeS{10.1145/2810103.2813687,10.1145/3540250.3549093,10.1145/3293882.3330580,secrypt19,9590404,9676691,10.1145/3510003.3510088},
\emph{transparency} \citeS{pmlr-v81-buolamwini18a,10.1145/3338906.3338937,10.1145/3368089.3409697,10.1145/3243734.3243792}, \emph{scalability} \citeS{8418593,huang2017safety,10.1145/3510003.3510051,10.1145/3533767.3534373},
\emph{accountability} \citeS{7961993,pmlr-v81-buolamwini18a,10.1145/3368089.3409697}, 
\emph{Capacity} \citeS{10.1145/3530019.3530025,10.1145/3506695,10.1145/3551349.3556906}, \emph{efficiency} \citeS{10.1145/3540250.3549102,Chen_2022_CVPR}, \emph{stability} \citeS{10.1145/3533767.3534373,10.1145/3510003.3510088}, \emph{replaceability} \citeS{10.1145/3510003.3510051,10.1145/3368089.3409668}, \emph{reusability} \citeS{10.1145/3510003.3510051,10.1145/3368089.3409668}, \emph{imperceptibility} \citeS{Eykholt_2018_CVPR}, \emph{ethics} \citeS{10.1145/3368089.3409697}, 
\emph{flexibility} \citeS{10.1145/3293882.3330580},  \emph{sustainability} \citeS{10.1145/3510003.3510051}, 
\emph{reproducibility} \citeS{10.1145/3510003.3510163}, and 
\emph{availability}\citeS{10.1145/3540250.3549102}, \emph{adaptability} \citeS{10.1145/3510003.3510088}.

Interestingly, our systematic exercise could identify not only non-functional requirements that emerged already in previous studies \cite{habibullah2021non,habibullah2022non,horkoff2019non}, but also additional categories that were not previously pointed out, i.e., \emph{transferability}, \emph{cost}, \emph{accountability}, \emph{energy consumption}, \emph{sustainability}, \emph{capacity}, \emph{stability}, \emph{behavioral}, \emph{imperceptibility}, \emph{efficiency}, \emph{availability} and \emph{adaptability}.

\begin{figure}
\includegraphics[width=1\linewidth]{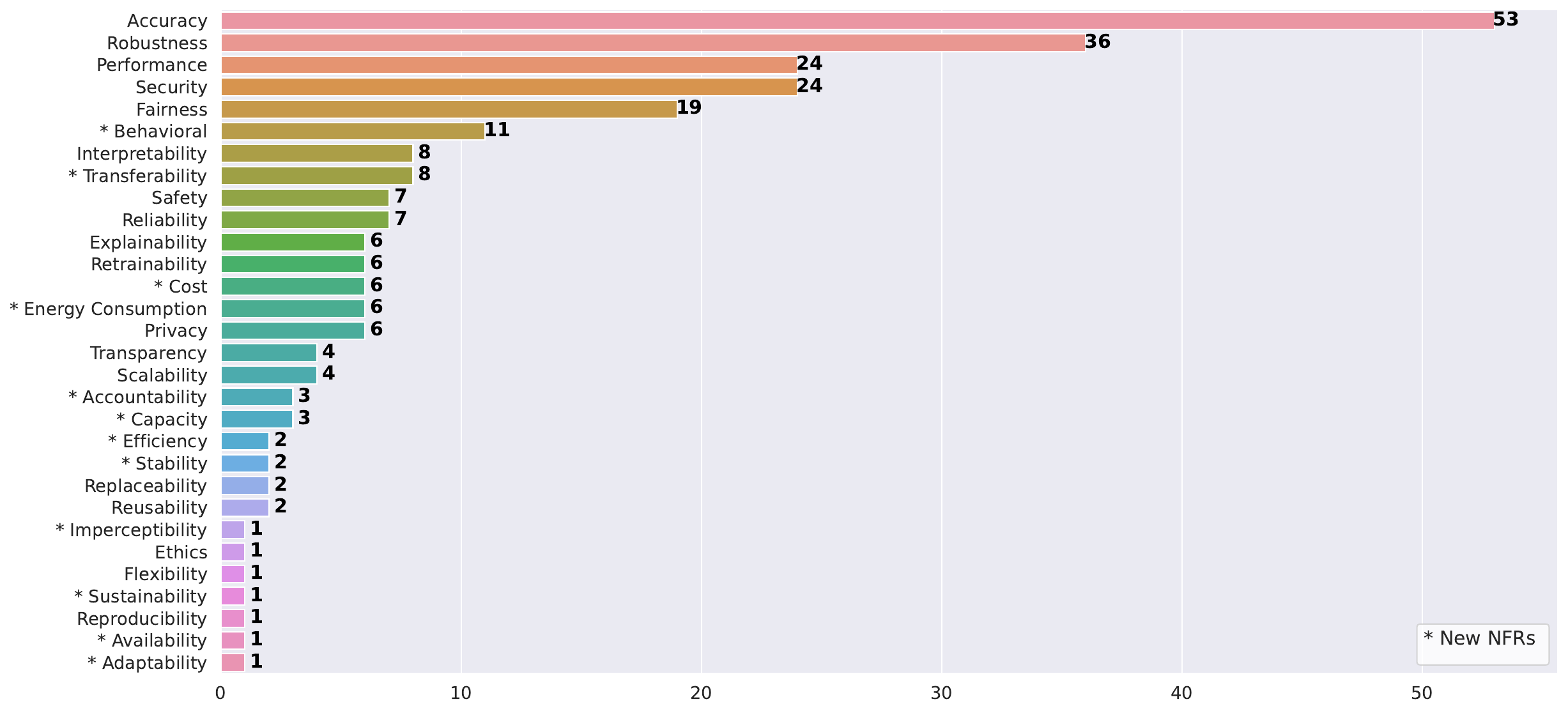}
\caption{Source-wise distribution of non-functional requirements in machine learning-enabled systems. A symbol `*' associated to one non-functional requirement denotes that it is a newly classified one.}
\label{fig:nfrs}
\end{figure}

Figure \ref{fig:nfrs} depicts the number of primary studies that targeted each of the 30 non-functional requirements identified. In the figure, the symbol `*' denotes the new non-functional requirements classified though our study, namely the ones that did not emerge from previous studies. As somehow expected, non-functional requirements such as \emph{accuracy} and \emph{robustness} of ML-enabled systems were those more frequently considered (53 and 36 times, respectively): these two aspects are likely the ones considered as recurrent and fundamental features of any machine learning model. Nonetheless, aspects connected to \emph{performance} (24), \emph{security} (24), \emph{fairness} (19), and \emph{behavioral (11)} seem to be growing research areas. On the contrary, multiple aspects were little explored, e.g., \emph{reusability} or \emph{scalability}, possibly indicating emerging research themes that would be worth further exploring. 

\begin{table}[ht]
    \caption{Non-Functional Requirements in Ml-enabled systems.}
    \label{table:list_nfrs}
    \centering
    \resizebox{\textwidth}{!}{    
    \tiny
    \rowcolors{1}{gray!15}{white}
    \begin{tabular}{|c|c|p{9cm}|}
    \hline
    \rowcolor{black}
    \textcolor{white}{\textbf{Cluster}} & \textcolor{white}{\textbf{NFRs}} & \textcolor{white}{\textbf{Definition}}\\
    Accuracy & Accuracy & The degree to which a model's predictions match the actual values.\\
    \hline
    Efficiency & Performance & The ability of a system to perform actions within defined time or throughput bounds.\\
        \cline{2-3}
        & Capacity & The amount of space required to store the model and any associated data.\\
        
        \cline{2-3}
        & Stability & Degree to which the output of a model varies as a consequence of perturbations to its input.\\
        
        \cline{2-3}
        & Scalability & The capability to handle increased workloads by adding resources while maintaining or enhancing model performance. \\
        
    \hline
    Maintainability & Replaceability & The degree to which a model can be replaced or substituted with another model without significant changes to the system.\\
        \cline{2-3}
        & Retrainability & The degree to which a model can be retrained on new data without significant performance loss.\\
        
        \cline{2-3}
        & Reproducibility & The degree to which a model's results can be reproduced by others using different software or hardware.\\
        
        \cline{2-3}
        & Transferability & The degree to which a model trained on one data set can be applied to another with similar characteristics.\\
        
        \cline{2-3}
        & Reusability & The degree to which a model can be reused in different applications or contexts.\\
        

        \cline{2-3}
        & Adaptability & The ability of the model to adapt to changing requirements or environments.\\
        
    \hline
    Resiliency & Security & The degree to which a model and its associated data are protected against unauthorized access, modification, or theft.\\
        \cline{2-3}
        & Safety & The degree to which a model and its outcomes are safe for humans and the environment.\\
        
        \cline{2-3}
        & Privacy & The degree to which a model and its associated data protect individuals' privacy rights and comply with data protection regulations.\\
        
        \cline{2-3}
        & Robustness & The ability of a model to maintain its performance when faced with uncertainties or adversarial conditions.\\
        
        \cline{2-3}
        & Reliability & Degree to which a model is resilient to errors and to variations of the surrounding environment.\\
        
        \cline{2-3}
        & Behavioral & The degree to which a model's outcomes align with requirements and expectations.\\
        
        \cline{2-3}
        & Flexibility & The degree to which a model can adapt to input data or environment changes without significant performance degradation.\\ 

        \cline{2-3}
        & Availability & The degree to which a model is operational and accessible when needed, without significant downtime or interruption. \\
    \hline
    Sustainability & Fairness & The degree to which a model produces unbiased predictions and decision-making outcomes across different groups of individuals.\\
        
        \cline{2-3}
        & Ethics & The degree to which a model mitigates potential societal risk.\\
        
        \cline{2-3}
        & Accountability & The degree to which individuals or organizations are held responsible for the actions of the model and its outcomes.\\
        
        \cline{2-3}
        & Cost & The overall economic means required to develop and maintain an ML-enabled system.\\
        
        \cline{2-3}
        & Energy Consumption & The amount of energy required for training and inference of the model and its impact on the systems.\\
        
    \hline
    Usability & Interpretability & The degree to which a model's predictions and decision-making process can be explained in terms of causality or human-understandable concepts.\\
        
        \cline{2-3}
        & Imperceptibility & The system's ability to produce outputs that are indistinguishable from what a human would produce in the same scenario.\\
        
        \cline{2-3}
        & Explainability & The degree to which a model's predictions can be explained and understood by humans.\\
        
        \cline{2-3}
        & Transparency & The degree to which a model's inner workings and decision-making process can be understood and evaluated by humans.\\
    \hline
    
    \end{tabular}
}
\end{table}


The classification exercise led to the identification of six main classes of non-functional requirements: Table \ref{table:list_nfrs} reports, for each class, the set of non-functional requirements belonging to the class, along with their definition. More specifically, we identified the following classes:

\begin{description}[leftmargin=0.3cm]

    \item[Accuracy.] The first class contained only the accuracy requirement, namely the degree to which a model's predictions match the actual values. We considered accuracy as a non-functional property for two main reasons. In the first place, multiple papers included in our systematic literature review, e.g., \cite{habibullah2021non,horkoff2019non,habibullah2022non}, described accuracy as a non-functional requirement, defining it as the number of correctly predicted data points out of all the data points. As our work aims at synthesizing the current knowledge available on the matter, we preferred to be conservative and considered accuracy as reported in the state of the art. In the second place, the definition is in line with the concept of non-functional requirement, i.e., \emph{``a condition that specifies a criterion that may be used to judge the operation of a system rather than specific behaviors''} \cite{bruegge2009object}: as a matter of fact, accuracy does not define constraints on the specific functionality that a system should enable, but rather question how the system performs in terms of correctness of the predictions made. In other terms, accuracy represents an attribute that can be used to judge the operation of a system rather than its specific behavior, i.e., it is by definition a non-functional requirement. The class was made isolated because of two main reasons: (1) accuracy represents the key feature to optimize by machine learning-enabled systems; (2) the non-functional requirement cannot be conceptually compared to any other, being ``unique''.

    \smallskip
    \item[Efficiency.] This was concerned with the overall performance and effectiveness of machine learning-enabled systems. The ISO/IEC 25010 standard includes performance efficiency as one of its main characteristics, encompassing attributes such as time behavior, resource utilization, and capacity. As such, we exploited the same reasoning to cluster non-functional requirements such as \emph{performance}, \emph{capacity}, \emph{stability}, and \emph{scalability}, under the \textsl{``Efficiency''} class.

    \smallskip
    \item[Maintainability.] Maintainability is chosen as a cluster to represent \emph{replaceability}, \emph{retrainability}, \emph{reproducibility}, \emph{transferability}, \emph{reusability}, and \emph{adaptability}: it emphasizes the ability of a system to adapt and evolve over time. All these non-functional requirements are critical to ensure that the system can be maintained and updated as needed and that it can be easily adapted to new use cases or environments. By prioritizing maintainability, designers and developers can create reliable, efficient, adaptable, flexible, and sustainable systems over the long term.
    


    \smallskip
    \item[Resiliency.] Resilience represents \emph{robustness}, \emph{reliability}, \emph{behavioral}, \emph{flexibility}, \emph{security}, \emph{safety}, \emph{privacy}, and \emph{availability}, as it emphasizes the ability of a system to adapt and recover from adverse events while maintaining its functionality and performance. Robustness, reliability, and behavioral flexibility are essential to ensure the system can operate in various conditions and remain responsive to changing circumstances. Security, reliability, and confidentiality are also critical to ensure the system protects users and stakeholders from harm, including threats to their physical safety, personal information, and data confidentiality. By prioritizing resilience, designers and developers can build reliable and secure systems that can withstand disruptions and threats while maintaining their core functions and services.
    
    \smallskip
    \item[Sustainability.] Sustainability is chosen as a cluster to represent economic, social, and environmental aspects. It ensures that resources are used efficiently and effectively, reducing the environmental impact, \emph{energy consumption} and financial \emph{costs} associated with producing the model and making inferences. Additionally, sustainability is chosen to represent \emph{fairness}, \emph{ethics}, and \emph{accountability}, as these aspects are critical for ensuring that the system benefits all stakeholders, including marginalized and vulnerable communities, and operates in a way that aligns with ethical principles and values. The rationale to classify social and ethical concerns under the \textsl{``Sustainability''} category comes from the increasing recognition that social sustainability is currently experiencing. For instance, the United Nations included the reduction of inequalities among the 17 objectives for sustainable development.\footnote{The United Nations Goals for Sustainable Development: \url{https://sdgs.un.org/goals}.} Researchers have been also arguing that fairness and ethical concerns should be considered as a form of sustainability. McGuire et al. \cite{mcguire2023sustainability} advocated that social sustainability refers to multiple dimensions, including pro-social vs. anti-social affordances. On a similar note, various other researchers \cite{lago2015framing,becker2015sustainability,dillard2008understanding} argued to consider social and ethical aspects as sustainability properties. Overall, sustainability can be considered as a guiding principle for designing and developing efficient, cost-effective but also equitable, ethical, and accountable systems.
    
    \smallskip
    \item[Usability.] The usability class is similar to the one defined by Habibullah et al. \cite{habibullah2022non}. More particularly, this class was chosen as the cluster to represent \emph{interpretability}, \emph{imperceptibility}, \emph{explainability}, and \emph{transparency}: it is concerned with ensuring that a system is easy to use and understand for end users, reducing the cognitive load required to interpret the model's behavior, and making it more transparent and explainable.
    


\end{description}

Researchers can use the taxonomy built in the context of our work to have a comprehensive mapping of the relevant non-functional requirements to optimize when developing machine learning-enabled systems, other than understanding the quality and efficiency aspects practitioners should focus on when releasing machine learning models.




\subsubsection{On the non-functional requirement domains}

\begin{figure}[h]
\includegraphics[width=0.9\linewidth]{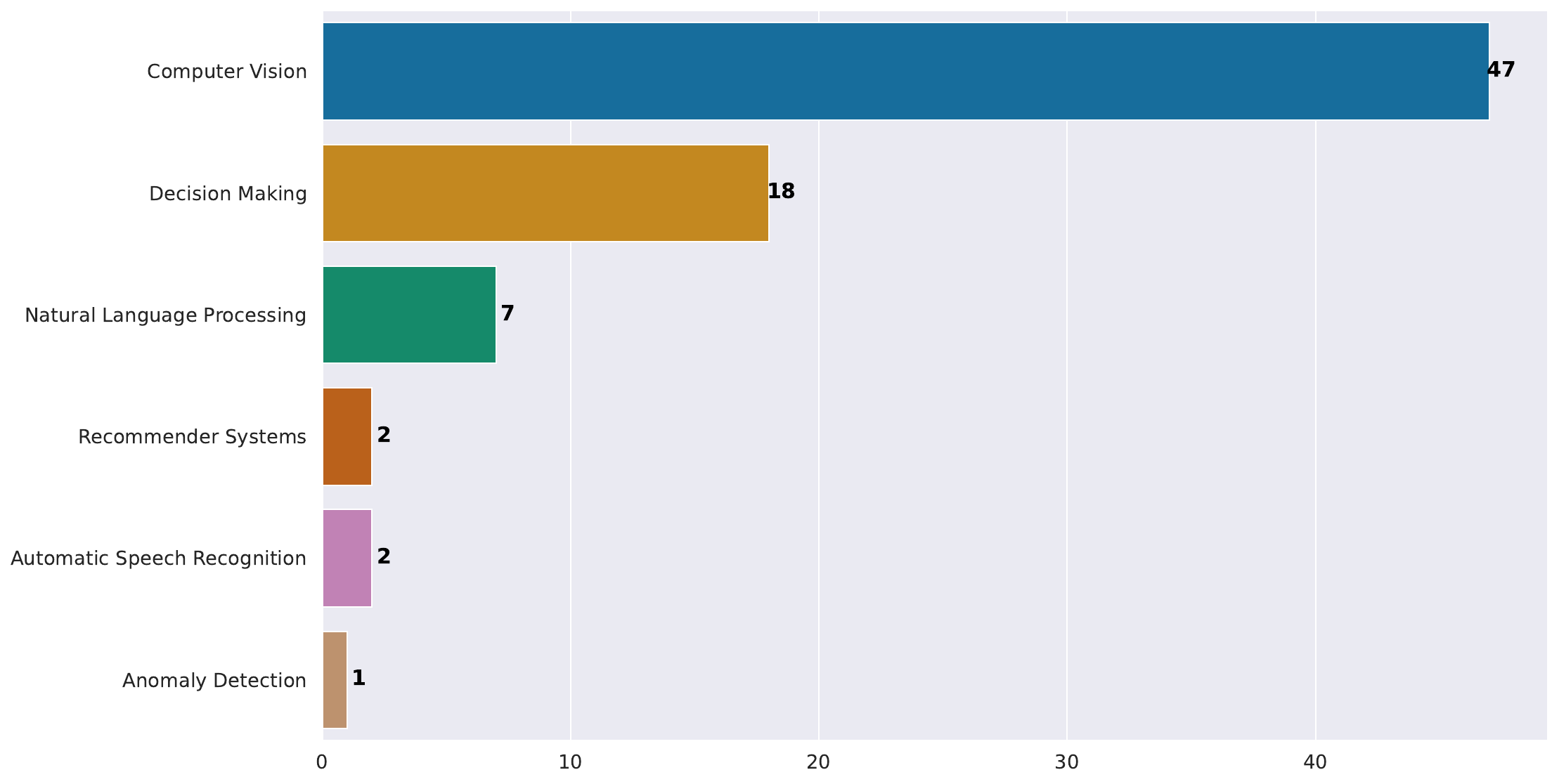}
    \caption{Distribution of NFRs across ML domains in ML-enabled systems.}
    \label{fig:domain_nfrs}
\end{figure}
\begin{figure}[h]
\includegraphics[width=0.9\linewidth]{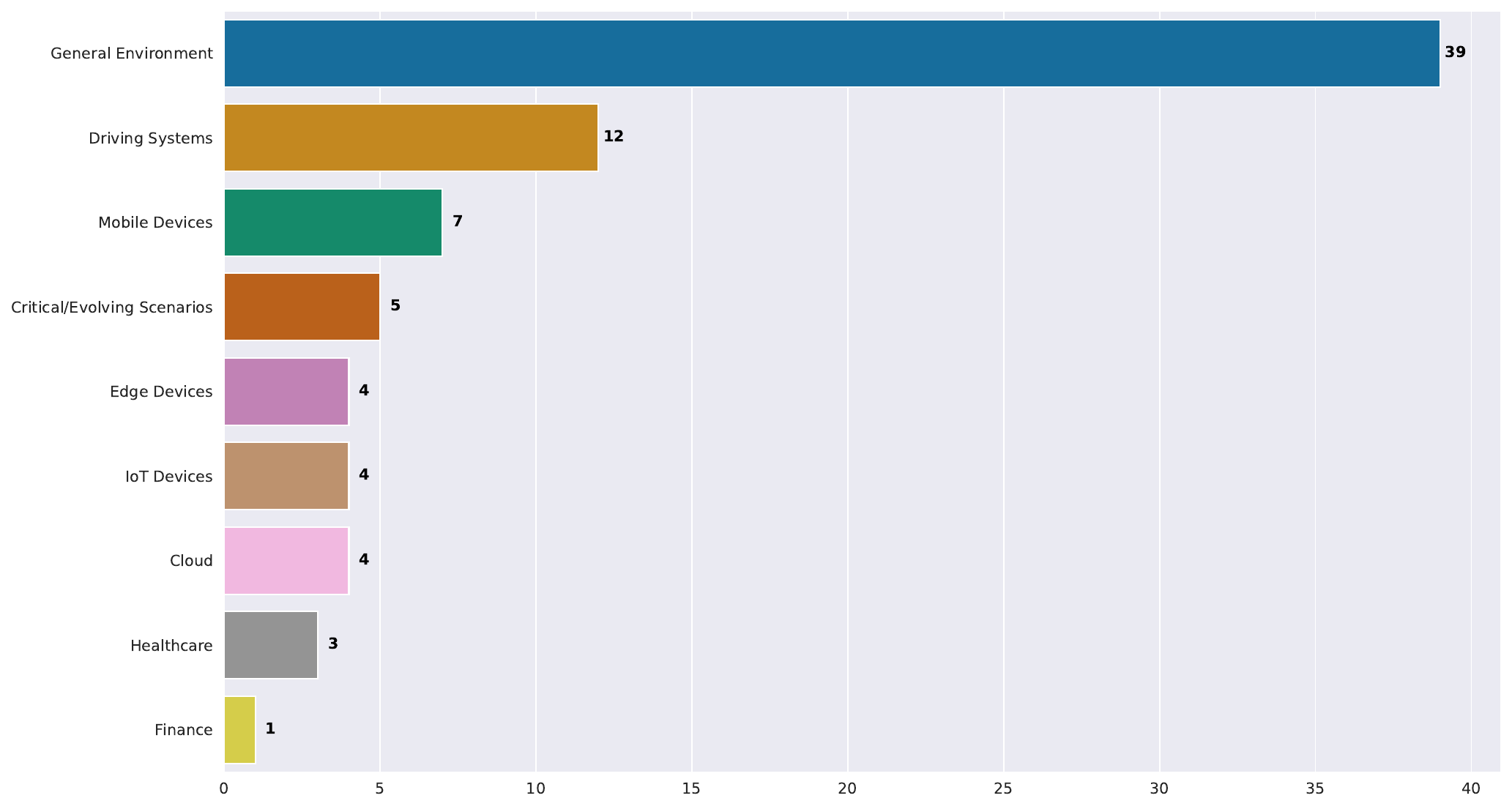}
    \caption{Distribution of NFRs across Environment domains in ML-enabled systems.}
    \label{fig:environment_nfrs}
\end{figure}

Figure \ref{fig:domain_nfrs} overviews the number of primary studies targeting each machine learning domain extracted. As shown, the most frequent are \textsl{`Computer Vision'} (47), \textsl{`Decision Making'} (18), and \textsl{`Natural Language Processing'} (9), while other, less targeted domains pertain to emerging technologies, e.g., \textsl{`Recommender Systems'} or \textsl{`Automatic Speech Recognition'}. These insights raise some key ML domains that researchers have been analyzing in the recent past, but also raise contexts where further research might be worth focusing on. The much larger amount of primary studies targeting \textsl{`Computer Vision'} may indicate the critical nature of such a domain with respect to the management of non-functional requirements. This may be due to the nature of the inputs that the machine learning-enable systems should consider, i.e., images or videos, which may be the subject of multiple concerns such as accuracy, security, ethics, and privacy---this may potentially emphasize the need for novel methods to process images and/or deal with non-functional attributes in the field.

\begin{figure}[h]
\includegraphics[width=1\linewidth]{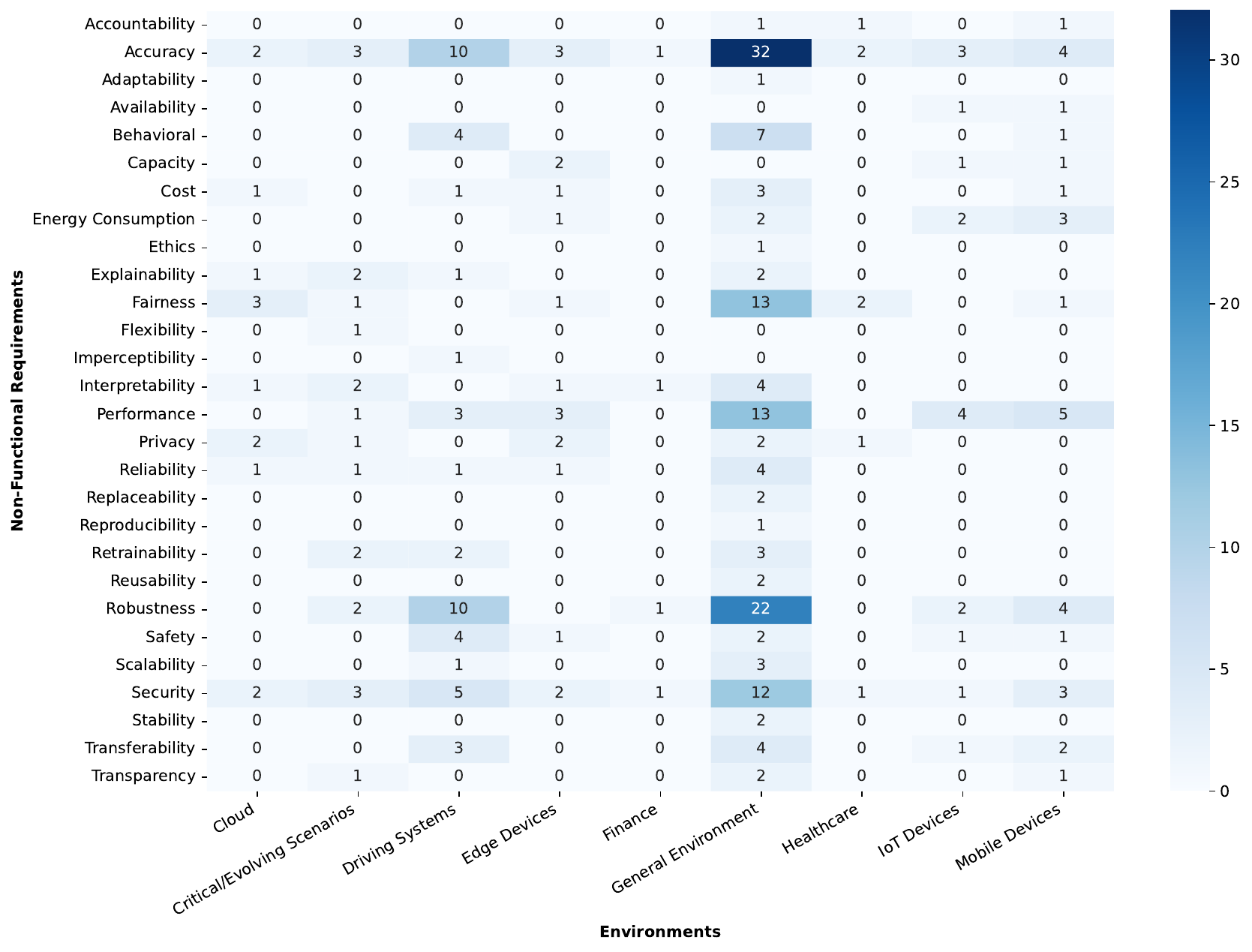}
\caption{Heatmap of Non-Functional Requirements across Environment domains in ML-enabled systems.}
\label{fig:heatmap_nfrs}
\end{figure}
Besides the ML domains, we also analyzed the environments where the primary studies focused on. These are shown in Figure \ref{fig:environment_nfrs}. The most common environment has been classified as \textsl{'General Environment'} (39) and indicates that most papers did not explicitly specify the environmental conditions which they experimented with. Other papers were instead more specific, targeting \textsl{'Driving Systems'} (12), \textsl{'Mobile Devices'} (7), and \textsl{'Critical/Evolving Scenarios'} (5). Looking at the figure, we may conclude that most primary studies worked in generalized conditions, with fewer researchers deepening the analysis of non-functional requirements in the context of ML-enabled systems designed to work under specific working conditions. 

This may represent a challenge for further research: researchers might indeed be interested in further understanding the peculiarities of the various environments, letting emerge novel non-functional requirements to deal with or learning how to prioritize non-functional requirements based on the environmental conditions that the ML-enabled systems are called to face.

Providing an additional perspective, Figure \ref{fig:heatmap_nfrs} illustrates a heatmap that shows the relation between the specific non-functional requirements presented in Table \ref{table:list_nfrs} and the corresponding environments depicted in Figure \ref{fig:environment_nfrs}. Each entry in the heatmap represents a numerical value, indicating the frequency with which the `i-th' non-functional requirement has been considered within the `j-th' environment, with `i' corresponding to the rows and `j' to the columns. By looking at the columns, the heatmap shows which are the non-functional requirements often addressed within the same environment. By looking at the rows, the heatmap shows the environments where each non-functional requirement has been analyzed so far. Through this visualization, we could first observe that there exist a number of environments where the current knowledge seems to be limited: for example, the environments concerned with \textsl{`Finance'}, \textsl{`IoT'}, and \textsl{`Cloud'}, and \textsl{`Edge Devices'} were the least targeted ones, which possibly suggests that further research might consider the impact of non-functional requirements in these contexts.

As expected, \emph{accuracy} represents the property more frequently addressed in all the environments. Nonetheless, some non-functional requirements like \emph{robustness}, \emph{performance}, \emph{fairness}, and \emph{security} have been targeted much more than others, possibly indicating that researchers perceived these as essential properties to investigate. At the same time, Figure \ref{fig:heatmap_nfrs} could further highlight the non-functional requirements that were somehow neglected so far: for instance, all the properties concerned with maintainability were found to be mostly unexplored independently from the environment considered. These insights may, again, be beneficial for researchers interested in working toward understanding non-functional requirements and their impact on ML-enabled systems working in different environments, and their understanding may help make more informed decisions and improve system development practices.

Finally, Figure \ref{fig:heatmap_nfrs} also suggests the existence of possible trade-offs to further consider in future research. Indeed, while the figure does not indicate if more non-functional requirements were simultaneously addressed by researchers, it may still highlight potential relations. As an example, in the \textsl{`Driving Systems'} environment we noticed that \emph{accuracy} and \emph{robustness} were more frequently targeted, possibly indicating that these two non-functional requirements should jointly considered by researchers to enlarge the current body of knowledge on the trade-offs among non-functional requirements of ML-enabled systems.


\rqanswer{1}{As a result of \textbf{RQ$_1$}, we could identify a total of 30 relevant non-functional attributes, classified according to six main categories. We also identified the most frequent ML domains and environments where these non-functional attributes have been investigated. The resulting taxonomy maps the current knowledge on the matter and provides insights into the research areas that may be worth exploring in the future.}




\subsection{\textbf{RQ$_2$} - What are the challenges of dealing with non-functional requirements of ML-enabled software?} 
The second research question is concerned with the identification of the challenges faced in dealing with the non-functional requirements of machine learning-enabled software. Taking the primary studies as an input, we processed them individually. We first went through the content of each primary study in search of two pieces of information: (1) the `context' where the non-functional requirement was considered; and (2) the `problem(s)' the primary study aimed at addressing. The information about the context was helpful to elicit where a non-functional requirement should be more carefully taken into account, while the information about the problem considered by the primary study was used to elaborate on the motivations making a non-functional requirement hard to treat. After collecting these pieces of information, we then applied an inductive reasoning to elicit the challenge(s) associated with the management of non-functional requirements. This process was systematic and included (1) the identification of the issues addressed by the primary studies; (2) the thematic analysis of these issues; and (3) the classification of the challenges related to management of a non-functional requirement. As such, the set of challenges described in our work has been derived from the content of the primary studies through the analysis of the contexts and problems considered by these primary studies. In doing so, we focused on the software engineering perspective, hence attempting to produce open challenges that our research community is called to address through further investigations.

Our work identified more than twenty software engineering challenges for our research community. First and foremost, it is worth pointing out that most of the challenges identified pertained to the use/integration of neural networks within more complex software systems. This was somehow expected, as deep learning has become the most widely employed form of machine learning to empower the capabilities of traditional software systems \cite{giray2021software}. In the second place, accuracy represents a cross-cutting concern, i.e., all the challenges identified have implications for the accuracy of machine learning-enabled systems. For this reason, we preferred to discuss accuracy while presenting the challenges connected to the other classes of non-functional requirements identified in \textbf{RQ$_1$}. The list of challenges is presented in the following.

\begin{description}[leftmargin=0.4cm]
    \item[\faSearch\ Efficiency (E):] In terms of overall performance and effectiveness of machine learning-enabled systems, we could elicit three major challenges which are connected to internal errors, high latency, and memory issues. Specifically:

    \smallskip
    \begin{itemize}
        \item \textbf{Challenge E.1 - Dealing with Internal Errors.} Neural networks can produce erroneous outputs as a consequence of internal errors due to \emph{incorrect parameters}, \emph{incorrect weight values}, \emph{uncovered root causes}, and \emph{incorrect manual labeling}. Those internal errors may significantly impact the performance of machine learning-enabled systems, leading to reduced accuracy, slower inference times, and increased energy consumption \citeS{9439863,10.1145/3530019.3530025,10.1145/3506695}. For example, incorrect weight values or parameters can cause the network to converge slowly or get stuck in suboptimal solutions, resulting in poor performance. Dealing with internal errors has two main connotations such as tracking errors in deep neural network models and addressing them. Our findings suggest that research in terms of \emph{bug localization}, \emph{performance monitoring}, and \emph{program repair} might support the challenges identified.
        
        \smallskip
        \item \textbf{Challenge E.2 - Dealing with High Latency.} Literature reported that multiple design choices related to \emph{hyper-parameter tuning}, \emph{model optimization}, and \emph{model architecture} can affect the overall efficiency of neural networks, even though those decisions might sometimes lead to increased accuracy \citeS{10.1145/3506695}. In other terms, the definition of neural networks should be considered as an optimization problem, where designers look for the best trade-off between performance and other quality measures and carry out preliminary studies to find the best configurations to employ within the specific use case. Additionally, the computational efficiency of DNN systems is very sensitive to even slightly different inputs. As a consequence, a change to the inputs may result in a significantly higher computational demand, deteriorating the overall models' efficiency \citeS{10.1145/3540250.3549102,10.1145/3551349.3561158,Chen_2022_CVPR}. 
        As such, researchers in the field of \emph{software analytics}, \emph{software architecture}, and \emph{software quality and optimization} might help address this challenge by means of analytical and code quality instruments aiming at finding the right compromise between accuracy and efficiency. 

        \smallskip
        \item \textbf{Challenge E.3 - Dealing with Space.} A third critical challenge arising from our review consists of the space required by machine learning models and, particularly, by neural network-based solutions. These may indeed present memory issues due to their large size, which has been particularly relevant in contexts like edge computing and IoT. Deploying large models on such devices can result in memory constraints affecting system performance \citeS{10.1145/3530019.3530025,10.1145/3506695,10.1145/3551349.3556906}. In response to this challenge, designers should be able to apply policies and strategies to build reduced models before deploying them on devices. These strategies may mitigate memory issues and improve the overall performance of machine learning-enabled systems. In this respect, our findings call for further research on the emerging field of \emph{tiny machine learning} \cite{lin2020mcunet}, that is, the combination of hardware, algorithms, and software that support on-device sensor data analytics at low power. While some initial effort has been made in the artificial intelligence research community (e.g., \cite{disabato2022tiny,siang2021anomaly}), we point out a lack of knowledge of the software engineering side of the matter. 
        

    \end{itemize}

    \smallskip
    The three challenges identified were discussed in all the domains requiring the definition of large models, i.e., \textsl{`Computer Vision'}, \textsl{`Driving Systems'}, \textsl{`Natural Language Processing'}, \textsl{`Mobile Devices'} and \textsl{`IoT Devices'}. This was somehow expected, as scalability concerns become evident in environments requiring high computational demands. 
    
    \smallskip
    \item[\faSearch\ Maintainability (M):] When it turns to the ability of machine learning-enabled systems to be adapted and evolved over time, we could elicit two main challenges connected to model reproducibility and model decomposition and reuse.

    \smallskip
    \begin{itemize}
        \item \textbf{Challenge M.1 - Increasing Model Reproducibility.} Machine learning-enabled systems based on deep learning models are known to be complex and difficult to reproduce accurately. One of the primary challenges in reproducing deep learning models concerns with randomness, which can affect the behavior of the algorithms used. Additionally, hardware non-determinism, such as that present in graphics processing units (GPUs), can further complicate the reproducibility of deep learning models \citeS{10.1145/3510003.3510163}. Our findings seem, therefore, to suggest the need for mechanisms that could support the \emph{verification} of deep learning models, for instance, by injecting elements of randomness and non-determinism to check how the system may behave in those circumstances.
        
        \smallskip
        \item \textbf{Challenge M.2 - Increasing Model Decomposition and Reuse.}
        A critical challenge when building and improving machine learning-enabled systems based on deep learning models is the need to reuse parts of previously constructed models. This can be difficult, for instance, when attempting to replace potentially defective parts with others.  A possible approach concerns with the identification of the parts responsible for each output class or module in the models: this would allow to selectively reuse or replace only the desired output classes to build the model. This challenge is strictly connected to the properties of randomness and non-determinism already discussed in the previous point. Indeed, increasing decomposition and reuse requires a proper understanding of the potential sources of variability and non-determinism in the system's software and hardware components. Therefore, improvements in terms of reproducibility may also lead to more reusable models \citeS{10.1145/3510003.3510163,10.1145/3510003.3510051,10.1145/3368089.3409668}. On the one hand, our findings reinforce the need for research in terms of \emph{verification}. On the other hand, the challenge of identifying reusable components seems to call for research in terms of \emph{program slicing} and \emph{refactoring}, which might support machine learning engineers to properly extract parts of the models for reusability purposes. 

    \end{itemize}

    \smallskip
    Challenges connected to maintainability were found to be pervasive in all domains, hence representing concerns that may impact any kind of machine learning-enabled system. 
    
    \smallskip
    \item[\faSearch\ Resiliency (R):]
        Improving the universal robustness and security of machine learning models present key issues \citeS{10.1145/3551349.3556920}. This was the class of non-functional requirements where we identified more challenges (8); these were connected to various aspects such as adversarial attacks, theft of model and intellectual property, attack and defense of models, repairing internal models, post-deploy issues, preserving privacy, and vulnerability transferability.

        \smallskip
        \begin{itemize}
        \item \textbf{Challenge R.1 - Resilience to Adversarial Attacks.} Machine learning models, and more particularly deep neural networks, are vulnerable to the so-called  \emph{adversarial attacks} \cite{https://doi.org/10.48550/arxiv.1706.06083,https://doi.org/10.48550/arxiv.1312.6199}, namely attacks that manipulate the input data in a way that causes the neural network to produce an incorrect output and significantly drop accuracy. Different types of adversarial attacks can be targeted or untargeted and can be carried out through various techniques such as \emph{adversarial examples} \cite{https://doi.org/10.48550/arxiv.1312.6199}, \emph{adversarial perturbation} \cite{NIPS2016_980ecd05}, \emph{poisoned data} \cite{10.1145/2046684.2046692}, and \emph{backdoor samples} \cite{8685687}. These types of attacks can negatively affect the overall robustness of the model and decrease its accuracy. As such, a key challenge is represented by the resilience to adversarial attacks \citeS{usman2021nn,7958570,9402124}. More specifically, existing literature highlighted the need for instruments able to design perturbations that may properly generate adversarial examples for different machine learning models, e.g., rotations, smoothing, and erosion operations in machine learning-enable systems targeting computer vision problems \citeS{9508369,huang2017safety,Eykholt_2018_CVPR,8746775}. Those instruments might support practitioners in terms of an improved comprehension of how models can be affected by vulnerabilities, other than through improved methods to increase appropriate defenses against adversarial attacks. Hence, our findings suggest additional \emph{verification and validation} research efforts: indeed, while the security perspective has been often targeted by the artificial intelligence community through the definition of new algorithms and techniques \cite{wang2019security,li2018security}, software engineering research might be complementary and propose novel verification and validation approaches, which may improve the policies and strategies employed by practitioners when testing for security. 

        \smallskip
        \item \textbf{Challenge R.2 - Exploiting the Sensibility of Adversarial Attacks.} A complementary perspective is concerned with the sensibility that certain adversarial attacks may have to input distortions and how such a sensibility can be exploited to improve the security of machine learning models. For example, in computer vision applications, attacks made through non-additive noise or geometric morphing may cause serious security concerns, yet those filters are sensitive to input distortions, meaning that automatic correction instruments may potentially distort the malicious inputs to mitigate their effects on the machine learning model \citeS{8418593,8746775,7958570,10.1145/3368089.3409739}. Our findings, therefore, point out the need for techniques that may handle perturbations that not only occur naturally in the physical environment but could be maliciously generated by adversary attacks. The perturbations are typically stealthy and undetectable, which poses further challenges for systems based on deep neural networks \citeS{9508369}: as such, characterizing them would be the first step for \emph{empirical software engineering} and \emph{automatic program repair} researchers. 
        
        \smallskip
        \item \textbf{Challenge R.3 - Security Verification of Pre-Trained Models.} Pre-trained models can be beneficial, as they reduce the computational burden of training complex deep learning models through transferability. However, pre-trained models may be vulnerable to attacks due to the displacement of the dataset and the potential for malicious code or exploitation \citeS{9766323,9402124}. On the one hand, assessing the security risks of these models still represent a key challenge to face through the definition of software analytics instruments able to provide practitioners with security insights and best practices in selecting the most appropriate pre-trained model to use to avoid security concerns. On the other hand, existing literature \citeS{9402124,9766323,9402020,10.1145/3510003.3510191} let also emerge additional challenges targeting the definition of design practices that may allow to develop and retrain security-aware fine-tuned models. Our findings, therefore, suggest that researchers in the field of \emph{software analytics}, \emph{software quality}, and \emph{verification and validation} might contribute to addressing the challenges in this set.
        
        
        \smallskip
        \item \textbf{Challenge R.4 - Resilience to Intellectual Property Theft.} Machine learning-enabled systems based on neural networks suffer from data breaches and unauthorized access to sensitive data due to training attacks \cite{10.1145/586110.586145}, i.e., malicious attacks on training data. According to the existing literature, only a little knowledge is available on how to secure machine learning models against those types of attacks. Greater attention should be paid to strengthening security protocols \citeS{secrypt19,9402124}. While this mostly calls the attention of researchers in the field of networks and security, the software engineering research community might still contribute through the definition of more security best practices able to protect models from intellectual property theft, especially in edge devices and IoT systems \cite{10.1145/3308558.3313591}. 


        \smallskip
        \item \textbf{Challenge R.5 - Diagnosing the Internal Behavior of Models.} Understanding the root cause of anomalous behaviors of deep neural networks and how to fix them represent critical aspects to further elaborate. On the one hand, evaluating the reliability implications of small sets of weights assigned to the network may support practitioners toward an improved understanding of how they work. In this respect, the definition of exploration, interpretability, and explainability methods may lead to substantial advances in terms of diagnosis of failures and misclassification, other than of security concerns  \citeS{10.1145/3460120.3484818,10.1145/3338906.3338937,10.1145/3180155.3180220,10.1145/3238147.3238187}---Ji et al. \citeS{10.1145/3533767.3534391} also called for novel local models to uncover the root causes of deep neural network failures. On the other hand, existing literature also pointed out the need for novel assessment metrics that, besides accuracy, can provide actionable insights into the overall robustness of the model \citeS{8952197,10.1145/3533767.3534373}. As such, our findings suggest multiple avenues for software engineering researchers in the field of \emph{fault localization}, \emph{explainable AI}, \emph{software metrics}, and \emph{automatic program repair}.
        
        \smallskip
        \item \textbf{Challenge R.6 - Optimal Post-Deployment Simulation.} While some threats to resiliency might be managed at the development time, additional challenges arise at the deployment stage. In particular, machine learning models could not work as expected when deployed. The problem has been mainly pointed out in the context of automatic speech recognition \citeS{10.1145/3533767.3534391,10.1145/3533767.3534408}, yet it may affect any kind of machine learning solution. The main challenge is related to properly understanding how the model would work in a real-case scenario. Besides software testing \cite{riccio2020testing}, literature identified post-deployment simulations as a complementary instrument that might be worth to further investigate: this refers to the definition of agent-based models that may (i) simulate the environment where the machine learning-enabled system would work and (ii) verify the system against a large variety of simulated inputs. This is, again, something that may catch the attention of researchers in the field of \emph{verification and validation} and \emph{machine learning for software engineering}.

        \smallskip
        \item \textbf{Challenge R.7 - Preserving Privacy in Machine Learning-Enabled Systems.} The primary studies identified the challenge of developing privacy-preserving deep neural networks and machine learning systems, which may allow algorithms to run securely on distributed data without compromising sensitive information about the subjects of the data, leaving users with the ability to delete their data at any time 
        leaving users with the ability to delete their data at any time. This would require the definition of strategies to train models without sacrificing accuracy while ensuring the protection of private data \citeS{9676691,10.1145/2810103.2813687}. As such, this challenge targets multiple software engineering fields, from \emph{software quality} to \emph{software architecture}, which may deepen the current knowledge on data encryption, anonymization techniques, federated learning, and differential privacy. 

        \smallskip
        \item \textbf{Challenge R.8 - Improving the Generalizability of Existing Automated Frameworks.} Multiple primary studies identified limitations and challenges related to the automated support provided by existing frameworks. Challenges in this category mainly concern with their generalizability. For instance, \textsc{Plum} \citeS{9623049}, a recommendation system to prioritize model repair strategies, is limited to the analysis of deep learning models and does not provide support for different learning tasks. Similarly, other approaches \citeS{10.1145/3510003.3510231,10.1145/3132747.3132785} can only assist practitioners when developing specific types of neural networks. As such, we also identified technological and empirical challenges for our research community, which is called to develop and experiment with automated solutions by keeping generalizability into account.
    \end{itemize}    

    \smallskip  
    Challenges in this category were mainly considered in the domains of \textsl{`Computer Vision'}, \textsl{`Driving Systems'}, \textsl{`Mobile Devices'}, \textsl{`Edge Devices'}, and \textsl{`IoT Devices'}, i.e., those more intensively threatened by security issues. Nonetheless, we observed a rising interest toward \textsl{`Automatic Speech Recognition'}---this is likely due to the growing research on large language models \cite{chen2021evaluating}.
    
    \smallskip
    \item[\faSearch\ Sustainability (S):] As for the challenges in this category, we could observe that most of the available literature focused on ethics and fairness, somehow neglecting other perspectives of sustainability. More particularly, we identified seven main challenges connected to algorithmic discrimination, model accountability, fairness metrics, low-quality datasets, energy cost, energy and performance aware trade-offs, and generalizability of existing solutions. 
    
        \begin{itemize}
        \item \textbf{Challenge S.1 - Dealing with Algorithmic Discrimination.} Improving performance, accuracy, and fairness simultaneously is currently the main challenge for researchers \citeS{10.1145/3540250.3549093,10.1145/3540250.3549103,10.1145/3540250.3549169,10.1145/3468264.3468537}. More specifically, the challenge is to study and mitigate the impact of disparate results, offensive labeling, and uneven algorithmic error rates in data-driven applications. In addition, the primary studies investigated highlighted the relevance of dealing with algorithmic discrimination in the context of the pre-processing phase in an effort to improve the trade-off between fairness and performance in ML software. While the software engineering research community has already contributed to addressing this challenge (e.g., \citeS{7961993,10.1145/3338906.3338937}), our findings suggest that further research in terms of \emph{software analytics} and \emph{verification and validation} might be worthwhile. 

        \smallskip
        \item \textbf{Challenge S.2 - Model Accountability:} The term `accountability' refers to the model's ability to provide clear explanations of its decisions, its transparency in how it has been trained and makes decisions, and its ability to allow users to provide feedback or challenge its decisions. The primary studies considered in our work remarked that more effort should be given to this aspect to allow models to be deployed with an eye on ethics and fairness. In this respect, major challenges pertain to the definition of \emph{verification and validation} instruments through which practitioners can verify how accountable their systems actually are \citeS{7961993,pmlr-v81-buolamwini18a}. For instance, decision making may represent ideal tools to assess the level of accountability of machine learning models.

        \smallskip
        \item \textbf{Challenge S.3 - Fairness Analytics.} Another critical challenge for social sustainability is related to fairness analytics, that is, the definition of metric toolkits that may support practitioners in diagnosing fairness at the development time. However, fairness should not be considered as a non-functional requirement per se, but rather be part of trade-off analyses. More specifically, existing literature called for novel metrics able to provide insights into the compromise between fairness and accuracy \citeS{9402057}, fairness and privacy \citeS{10.1145/2810103.2813687}, and fairness and accountability \citeS{7961993}. As such, the challenge is represented by the diagnosis and multi-objective optimization of data-driven applications in multiple development stages, from requirements engineering to verification and validation \citeS{7961993}.

        \smallskip
        \item \textbf{Challenge S.4 - Improving Sustainability Benchmarks.} Current literature \citeS{pmlr-v81-buolamwini18a} also highlighted the lack of low-quality of benchmark datasets that may assist the creation of sustainable machine-learning applications. These datasets should indeed be devised by enabling cross-domain and cross-sectional analysis. On the one hand, application domains focusing on the image processing would require high-quality datasets allowing automated pose, illumination, and expression (PIE) analysis. On the other hand, datasets presenting large and diverse information on minorities should be devised to allow the more effective training of machine learning systems.
        
        \smallskip
        \item \textbf{Challenge S.5 - Reducing Energy Cost.} One of the most important challenges for deep neural networks is concerned with the reduction of energy and computational costs. Indeed, those systems have considerable energy and financial costs, which lead to increase CO2 emissions and memory consumption in both the training and post-release phases \citeS{10.1145/3506695,9590404,10.1145/3510003.3510231,10.1145/3510003.3510221,10.1145/3510003.3510088,10.1145/3551349.3561158}. More specifically, energy consumption issues significantly impact edge and IoT devices, which indeed represent critical use cases. The software engineering research community is called to contribute in different manners. First, \emph{empirical software engineering} research is called to assess the impact of different AI containerization strategies on energy consumption and memory utilization. Second, \emph{software quality} and \emph{software architecture} researchers might help address the trade-offs between training accuracy and post-deployment consumption. Last but not least, our findings call for further research on the definition of software engineering practices to develop \emph{tiny machine learning} solutions. 

        \smallskip
        \item \textbf{Challenge S.6 - Increasing the Practitioner's Awareness of Sustainability.} Our literature work identified the challenge of increasing the developer's awareness with respect to sustainability concerns. This issue is particularly relevant when considering energy consumption, which is a critical concern in today's world. As such, the definition of novel strategies and instruments to make practitioners aware of the trade-offs between performance and sustainability represents a key challenge for our research community \citeS{10.1145/3510003.3510221,10.1145/3530019.3530025}. Once again, more research on tiny machine learning might represent a suitable solution, as it would allow to improve the scalability of models without compromising their overall accuracy. At the same time, the \emph{empirical software engineering} research community might play a key role by understanding the practitioners' needs and practices, other than tailoring (novel) automated approaches on them. 

        \smallskip
        \item \textbf{Challenge S.7 - Improving the Generalizability of Existing Automated Approaches.} The last challenge in this category concerns with the generalizability of existing approaches that support sustainability analysis of machine learning-enabled systems. Multiple primary studies identified limitations and challenges that the research community should further consider. Similarly to \textbf{R.8}, current approaches can only provide support for the development of a limited set of deep neural networks \citeS{10.1145/3510003.3510202,10.1145/3510003.3510087}, are built using limited datasets \citeS{10.1145/3368089.3409697}, or are not integrated within MLOps pipelines \citeS{10.1145/3468264.3468565}. In other terms, the challenge emphasizes the need for novel, more accurate, generalizable, and integrated instruments that may support practitioners throughout the software lifecycle. 
        
    \end{itemize}

    \smallskip
    Since most of the challenges described pertained to social sustainability, these were investigated in the \textsl{`Decision Making'} and \textsl{`Healthcare'} domains, which are the most demanding in terms of ethics and fairness. However, we observed a growing interest in the \textsl{`Cloud'}, \textsl{`IoT Devices'}, and \textsl{`Edge Devices'} domains, mostly due to the increasing efforts on energy consumption matters. 
    
    \smallskip
    \item[\faSearch\ Usability (U):] Usability is the non-functional requirement with fewer challenges. Only a few works targeted this matter, hence suggesting that further research might be worthwhile. Indeed, usability, interpretability, and visualization of both shallow and deep learning models pose significant challenges when it turns to assess quality assurance throughout the software lifecycle. For instance, the lack of interpretability increases the effort required to estimate the soundness of machine learning-enabled systems. Current literature somehow neglected this research angle, especially when considering models operating in critical and evolving scenarios. As such, our findings suggest the urgent need for further research on usability concerns \citeS{9451178,10.1145/3338906.3338954,10.1145/3243734.3243792,10.1145/3429444,10.1145/3506695,10.1145/3447548.3467177}. Interestingly, the available primary studies did not focus on any specific application domain and proposed general-purpose interpretability models. This further suggests the investigation of domain-specific approaches and methods. 

    \smallskip
    \item[\faSearch\ Cross-cutting challenges (C):] Besides the challenges pertaining to the individual classes of non-functional requirements, we also identified a set of cross-cutting challenges which are related to the management and assessment of non-functional requirements.  
    
    \begin{itemize}
        \smallskip
        \item \textbf{Challenge C.1 - Context-based Trade-off Identification.} Finding a trade-off between multiple non-functional requirements represents a key socio-technical and managerial challenge \citeS{10.1145/3540250.3549093,10.1145/3510003.3510163,10.1145/3510003.3510163, 10.1145/2810103.2813687,10.1145/3510003.3510087}. On the one hand, machine learning-enabled systems should always preserve accuracy. On the other hand, other non-functional requirements may play a key role depending on the specific context where the system should be employed. As an example, robustness and security might be of paramount importance in the context of driving systems, while fairness might be preferred when dealing with healthcare systems. The identification of the non-functional requirements to preserve may indeed depend on contextual analyses performed at both socio-technical and managerial levels. In the former, novel approaches able to (semi-)automatically mine contextual information that provide practitioners with insights into the non-functional requirements to consider in the specific context might be worthwhile. In the latter, novel software project management methods to assess the significance of non-functional requirements might support the managerial decisions on how to design a machine learning-enabled system. This challenge therefore emphasizes the need for novel approaches to identify the non-functional requirements to simultaneously consider within a given context.

        \smallskip
        \item \textbf{Challenge C.2 - Prioritizing and Balancing Non-Functional Requirements.} A second cross-cutting challenge pertains to the balancing of multiple non-functional requirements. While the trade-off identification may provide practitioners with information on the aspects to take into account while developing a ML-enabled system, this would not be enough to prioritize and balance them. Literature argues the need for novel methods to support the technical management of non-functional requirements in terms of prioritization and balancing of non-functional requirements during the optimization process \citeS{10.1145/3510003.3510221,10.1145/3530019.3530025,10.1145/3540250.3549093, 10.1145/3338906.3338954,10.1145/3540250.3549103,10.1145/3551349.3556906}. As such, the \emph{requirements engineering} research community might further explore this matter, by proposing guidelines or automated instruments to (1) prioritize non-functional requirements and (2) find an optimal balance among both correlated and contrasting objectives by means of the analysis of contextual factors and practitioners' preferences.

        \smallskip
        \item \textbf{Challenge C.3 - Software Analytics for Non-Functional Requirement Assessment.} As a last challenge of this category, multiple primary studies advocated the need to empower requirements engineering processes with software analytics instruments able to assess the implications that trade-off choices may have on the development of ML-enabled systems \citeS{10.1145/3510003.3510088,10.1145/3551349.3561158,10.1145/3340531.3411980,10.1145/3447548.3467177,10.1145/3534678.3539145}. Current literature argues the need for software metrics able to inform practitioners of how their requirements engineering decisions may influence both the complexity of the development process and the overall quality of the system being developed. As an example, the definition of strategies to recommend the quality assurance mechanisms to put in place based on the trade-off to meet would be desirable \citeS{10.1145/3534678.3539145}. On a similar note, researchers claimed that the definition of novel predictive analytics instruments might largely improve requirements engineering processes by enhancing the practitioner's capabilities to assess the impact that trade-off analysis may have on the ML-enabled system throughout the software lifecycle. As a consequence, this challenge may be of the interest of \emph{requirements engineering} and \emph{software analytics} research communities, which may jointly collaborate toward an improved understanding of the support required by requirements engineers in practice.
        
    \end{itemize}
    
\end{description}

\rqanswer{2}{We elicited more than 20 software engineering challenges targeting both the individual categories of non-functional requirements identified in our systematic synthesis work and the cross-cutting aspects affecting non-functional requirements engineering as a whole. We methodically untangled the challenges plaguing machine learning-enabled systems, providing insights into the specific research fields interested in those challenges. According to our results, we call for comprehensive analyses and approaches that may assess the impact and implications of the management of individual and multiple non-functional requirements.}

%% file: sections/5_Discussions.tex
\section{Discussion and Implications}
\label{sec:discussion}
The results coming from our study provide a number of additional discussion points and implications that we further discuss in the following. 

\begin{description}[leftmargin=0.3cm]
	
    \item[On the Relation between the acquired knowledge and the previous one.] Our results allowed us to discover multiple aspects concerned with non-functional requirements of machine learning-enabled systems that extend the existing knowledge. The outcome of \textbf{RQ$_1$} provided a comprehensive classification of non-functional requirements along with their classes. Our work extended the insights provided by Habibullah et al.  \cite{habibullah2022non}, providing a larger overview of the issues that practitioners should consider while developing machine learning-enabled systems. Specifically, we noticed three major differences. First, Habibullah et al. \cite{habibullah2022non} did not include \textsl{`Sustainability'} as a category of non-functional requirements but, rather, they limited the categorization to aspects connected to ethics and bias. Hence, our work extends the classification by contextualizing the aspects considered by Habibullah et al. \cite{habibullah2022non} within the broader context of sustainability, explicitly referring to three dimensions such as environmental, economic, and social sustainability. Secondly, we identified four additional non-functional requirements connected to \textsl{`Resiliency'}: \emph{robustness}, \emph{reliability}, \emph{behavioral}, and \emph{flexibility}. In this sense, we were able to enlarge the knowledge of the non-functional requirements of ML-enabled systems, building on top of the existing classification schemas to further enrich them with additional characteristics and attributes that should be optimized. Finally, we also noticed a substantial difference in the treatment of the category \textsl{`Maintainability'}. Habibullah et al. \cite{habibullah2022non} combined the traditional concept of maintainability, i.e., the ability of a system to be modified, improved, and adapted, with the concept of usability, i.e., how effectively a user can learn and use the system. Based on our systematic work, we found out that the concepts would be better split into two separate categories. The rationale is twofold. On the one hand, maintainability and usability can be logically considered as two different non-functional requirements which pertain to distinct desirable attributes of ML-enabled systems. On the other hand, the optimization of these two attributes would require different methods and strategies and, for this reason, further research would benefit from a clear distinction between the two. The additional knowledge gathered by our work may serve as input to both researchers and practitioners. The former might take our review as a means to design empirical experimentation and approaches able to deal with the non-functional aspects that previous research did not consider. The former might instead take our classification as an input to increase their own awareness with respect to the problems arising when developing machine learning-enabled systems, other than reason on how to address those aspects in real-world use cases.
    
    \implication{1}{Our classification of non-functional requirements may inspire further research on the matter, other than making practitioners aware of the multiple, multi-faceted concerns arising when developing machine learning-enabled software systems.}
    
    At the same time, it is also worth commenting on how the findings coming from \textbf{RQ$_2$} inform future research. Our research identified a set of over 20 software engineering challenges that should be further considered by our community. When comparing our findings with those reported by Horkoff \cite{horkoff2019non}, we may first observe that the level of granularity of the challenges reported is, in general, different. Indeed, Horkoff \cite{horkoff2019non} discussed a set of \emph{conceptual issues} that may threaten the researchers' capabilities of effectively dealing with non-functional requirements of machine learning-enabled systems. For instance, the author reported that the understanding of the research community with respect to non-functional requirements of machine learning-enabled systems is fragmented and incomplete, which includes the lack of definitions of the key attributes affecting the way machine learning-enabled systems operate. On the one hand, our work was able to address some of the conceptual concerns raised by Horkoff \cite{horkoff2019non}, offering a theoretical framework describing the definitions of non-functional requirements, how they have been explored so far, and what are the potential inter-relations among them. On the other hand, the set of challenges identified may be seen as a more concrete instantiation of the conceptual concerns reported by Horkoff \cite{horkoff2019non}. Indeed, we were able to describe the specific challenges to address for each non-functional requirement, hence providing researchers with an actionable instrument to design further studies and investigations into the matter. In addition to that, it is also worth reporting that we could also corroborate one of the challenges identified by Horkoff \cite{horkoff2019non}, namely the need for engineering machine learning-enabled systems for reducing internal errors and security. So, in conclusion, we could elaborate a larger, comprehensive catalog of challenges that have to do with multiple aspects of SE4AI and that naturally inform a number of software engineering research communities, from the one on requirements engineering till the one on software testing. As such, our work can be considered as a cross-cutting knowledge base for serving further research in software engineering. 
    
    \implication{2}{Our work could identify a number of software engineering challenges that are not yet targeted by automated approaches, hence suggesting the future avenues that researchers can consider to further enlarge the knowledge and support provided to handle non-functional requirements of machine learning-enabled systems.}
     
    As a last point of discussion, we comment on how we envision our results have influence on the state of the practice. The outcome of \textbf{RQ$_1$} may possibly be exploited to support the definition of standards and guidelines that attempt to describe how to deal with non-functional requirements of machine learning-enabled systems. For example, the definitions provided might be potentially useful to support and complement an existing, emerging standard such as the \textsl{ISO/IEC 25059},\footnote{The \textsl{ISO/IEC 25059} standard: \url{https://iso25000.com/index.php/en/iso-25000-standards/iso-25059}} which defines an initial taxonomy of non-functional properties relevant for machine learning-enabled systems. In this respect, we foresee improved implementations of the standard, which takes our findings into account, hence enlarging the definitions available for researchers and practitioners. Furthermore, when comparing our findings against the research that explored the current state of the industrial practice, we see some interesting complementarities. In particular, Amershi et al. \cite{amershi2019software} conducted a case study in \textsc{Microsoft} where they let emerge a catalog of best practices to make software engineering actionable for the development of machine learning-enabled systems. Among these practices, the authors pointed out the need for model debugging and interpretability and for model evolution, evaluation, and deployment. In the first place, the outcome of \textbf{RQ$_1$} has the potential to make the best practices identified by Amershi et al. \cite{amershi2019software} more tangible: indeed, our findings provide a set of factors that strictly relate to the model debugging, interpretability, evolution, evaluation, and deployment which are potentially measurable and, therefore monitorable throughout the different development stages of machine learning-enabled systems. As an example, the \emph{retrainability} attribute pertaining to the \textsl{`Maintainability'} cluster may impact the evolution and deployment of machine learning-enabled systems. Our systematic synthesis provides a theoretical foundation to build novel measurement and monitoring systems that may be employed to produce higher-quality machine learning-enabled systems. In addition, the findings of \textbf{RQ$_2$} provide insights into the research gaps making the best practices identified by Amershi et al. \cite{amershi2019software} difficult to actually implement. Also in this case, our systematic synthesis may therefore provide researchers with an improved understanding of the next steps to pursue to better support the development and evolution of machine learning-enabled systems. 
    
    \implication{3}{Our findings complement the current state of the practice in different manners. First, the outcome of \textbf{RQ$_1$} may be useful to extend and/or complement emerging standards describing relevant non-functional requirements of machine learning-enabled systems. Second, the results of \textbf{RQ$_1$} and \textbf{RQ$_2$} can be combined with the pieces of information that emerged from the state of the practice, highlighting the current research gaps that should be filled and the potential opportunities of technological transfer.}

	\item[On the Inter-Relation among Non-Functional Requirements.] According to our results, and specifically, those coming from \textbf{RQ$_2$}, there exist some inter-relations among non-functional requirements. These are not only concerned with the relation that accuracy has with other requirements but also with the innate interconnections between multiple non-functional aspects playing a role in the development of machine learning-enabled systems. In this respect, researchers in the area of requirements engineering and empirical software engineering might cooperate toward the development of novel taxonomies that may map the relations among the non-functional requirements of machine learning-enabled systems and how they impact each other. Our results also called for further research on the identification, management, and assessment of the trade-offs among multiple non-functional requirements. 
	
	\implication{4}{Our findings suggest novel interconnections and research on the multi-objective optimization of non-functional requirements of machine learning-enabled systems. We call for a brand new research field focusing on empirically investigating the relations between non-functional requirements and how those relations may inform the development of automated approaches to optimize machine learning-enabled systems.}
	
	\item[A Managerial Viewpoint.] As a follow-up discussion, the research questions targeted by our work raise multiple trade-offs to consider when developing machine learning-enabled systems. The accuracy of artificial intelligence algorithms must necessarily be balanced with other aspects to create trustworthy software systems. On the one hand, this clearly represents a call for further research. On the other hand, our work has implications on the managerial side: the constant and continuous search for trade-offs indeed represents a high-level challenges for project managers, who are required to monitor and handle multiple non-functional attributes throughout the evolution of machine learning-enabled systems. In the first place, the results produced by \textbf{RQ$_1$} may allow the reader to understand what the set of non-functional attributes actually takes into account. Secondly, the challenges identified in the context of \textbf{RQ$_2$} should not only be considered from a technical perspective but also from a socio-technical and managerial one. Our work indeed calls for further research on the managerial strategies that would indicate the most appropriate management policies to apply when handling non-functional requirements. Also, we bring to the attention of software engineering and software project management researchers the lack of optimization approaches that may be contextual, dependent, and adaptable to the evolution of the system---we argue that those properties would be key to enabling the proper management of non-functional requirements over the software lifecycle. 

	\implication{5}{We call for research on managerial and socio-technical strategies to handle non-functional requirements throughout the evolution of machine learning-enabled systems. At the same time, our findings point out the need for optimization approaches specifically tailored to software evolution, hence being contextual, dependent, and adaptable.}
	
	\item[Machine Learning Sustainability: Software Engineering Comes to Rescue.] The results of \textbf{RQ$_2$} identified sustainability as a non-functional requirement that has been somehow neglected by our research community. While previous work has started exploring the social side of sustainability, we could identify a lack of software engineering investigations into energy- and cost-related concerns. On the one hand, our work identified \emph{tiny machine learning} \cite{lin2020mcunet} as a promising solution to some of the key challenges affecting the energy consumption of machine learning-enabled systems: as such, we call for additional investigations into these aspects. On the other hand, we also highlight a lack of experimentation in terms of costs: in this respect, further analysis of how to predict and manage costs, other than how to optimize non-functional approaches by taking those aspects into consideration, would be worthwhile. 
	
	\implication{5}{Sustainability concerns should be more carefully considered by the software engineering research community. There is a need for approaches to deal with energy consumption other than cost-aware non-functional requirements optimization strategies.}

\end{description}

%% file: sections/6_Threat.tex
\section{Threats to Validity}
\label{sec:threat}
The validity of both the research method and conclusions drawn in our systematic literature review might have been threatened by some key considerations. This section overviews the potential limitations of our study and how these were mitigated when designing the study.

\smallskip
\textbf{Literature selection.} \revised{A critical challenge for a systematic literature review is identifying a consistent and comprehensive body of knowledge concerning the subject of interest. In this respect, we first approached the search process by adhering to well-established guidelines \cite{kitchenham2009systematic}. Nonetheless, we realized that these guidelines would not have been sufficient to extract the required body of knowledge and, as a consequence, might have led to missing key resources. To mitigate this threat, we, therefore, opted for a \emph{hybrid} systematic analysis \cite{mourao2020performance}: we complemented the set of primary studies identified through the guidelines by Kitchenham et al. \cite{kitchenham2009systematic} with additional resources coming from (1) the systematic screening of the research articles published in top-tier software engineering and artificial intelligence venues \cite{wohlin2020guidelines}; and (2) multiple snowballing iterations of the incoming and outcoming references of the primary studies identified \cite{wohlin2016second}. In this respect, it is worth further discussing the seed search procedure conducted on artificial intelligence venues. As explained in Section \ref{seed_set}, we had to limit ourselves to the analysis of a subset of all the top-tier conferences and journals in the field of artificial intelligence to make systematic scanning sustainable and feasible. In doing so, we solely considered the venues having a higher likelihood to include engineering or empirical pieces of work that might have been relevant for addressing our research questions: among 35,155 papers published within the selected venues between 2012 and 2022, we only found 12 potentially relevant articles, which reduced to seven after the quality assessment stage. We are aware that an extensive analysis of the artificial intelligence venues might have identified some additional relevant primary studies: this remains a limitation of our work that the reader must be aware of. However, our analyses revealed that the amount of papers published in artificial intelligence venues that describe or optimize non-functional requirements is somehow limited (only 0.02\% of the papers scanned were finally included), possibly not justifying the effort that would have been required. In this sense, alternative research instruments, e.g., qualitative surveys or interviews with machine learning engineers, might be more useful to complement our findings and identify additional concerns or approaches used to optimize non-functional requirements.}

In the second place, it is worth remarking that we performed the search on multiple databases such as \textsl{ACM Digital Library}, \textsl{Scopus}, and \textsl{IEEEXplore}. This was done to ensure wider coverage of the primary studies published in the literature. 

The reliance on a hybrid systematic literature review, other than the multiple actions conducted to extend the search process, makes us confident of the completeness of the literature selection. However, for the sake of verifiability and replicability, our online appendix \cite{DeMartino2023} contains all the data and material used to produce each intermediate search of our study. The interested reader might use the material to either assess the soundness/completeness of the process and further build on top of our results \cite{wohlin2020guidelines}.

\smallskip
\textbf{Literature analysis and synthesis.} Upon completing the search process, we applied several steps to ensure the inclusion of the relevant primary studies to address the research questions driving our systematic literature review. In the first place, we designed a set of exclusion criteria with the intent of filtering out the primary studies that did not fit our scope. In addition, we proceeded with the definition of inclusion criteria, other than a formal quality assessment of the suitability of each identified resource for the goals of our study. 

These steps were conducted manually; therefore, the risk of subjectiveness and human error were the main limitations. While the first author mainly conducted all the activities, there are two considerations to make. In the first place, we designed the systematic literature review so that validation was conducted at the end of each step. The validation first involved the second author of the paper, who took the role of inspector: he was indeed called to verify the actions conducted by the first author and identify potential errors. At the end of this process, both authors opened a discussion and addressed the concerns raised in the validation. 

Secondly, the two authors constantly worked together to define the research method and actions to make the study sound and reliable. In particular, they met weekly to discuss the advances of the systematic literature review and the potential limitations to address. These weekly meetings further limited potential concerns due to subjectiveness and human bias. Nonetheless, we acknowledge that the entire process has been conducted based on the knowledge and expertise of the authors, which might be limited. On the one hand, we released all the material produced to be as transparent as possible \cite{DeMartino2023}. On the other hand, we can only report some background information - this may be useful to assess the study's validity and estimate the resources required to replicate our work. The two authors have a research experience of one and ten years, respectively. Both conduct or have already conducted quantitative and qualitative studies in the past, other than systematic literature/mapping studies on themes connected to software engineering for artificial intelligence, optimization of non-functional requirements, and software maintenance and evolution. Furthermore, they are both involved in the academic courses of \textsl{Software Engineering}, \textsl{Fundamentals of Artificial Intelligence}, and \textsl{Software Engineering for Artificial Intelligence} at the University of Salerno (Italy) - the second author is the lecturer of those courses, while the first is a teaching assistant. 

\revised{The data analysis of the primary studies allowed us to address the research questions of the study, providing a picture of the current state of the art, synthesizing the research efforts conducted so far, identifying the areas that are still neglected or unexplored, and pointing out the next research directions that the software engineering research community should pursue to better assist practitioners in their activities. In doing so, we could analyze a bit more than two primary studies for each non-functional requirement, on average (69 articles concerned with 30 non-functional requirements). While this might indicate the limited generalizability of our findings, we do not see this point as a real limitation: our systematic review has indeed an intrinsic value which is independent of the number of primary resources finally considered. We are aware that future research may classify additional non-functional requirements or identify further challenges which are not documented in this paper: our ultimate goal is exactly that of stimulating more research on the matter by providing a foundational basis describing the current body of knowledge and the limitations that should be addressed in the future.}

%% file: sections/7_Conclusions.tex
\section{Conclusion}
\label{sec:conclusion}
In this paper, we conducted a hybrid systematic literature review on non-functional requirements of machine learning-enabled systems---this was motivated by the increasing number of papers published on these aspects during the last ten years. We tackled two main research angles such as (1) the classification of the non-functional requirements investigated so far, and (2) the challenges to face when dealing with them. The results of the systematic literature review do not only summarized the current knowledge on the matter, but also opened new horizons and challenges that our research community should consider. It is our hope that researchers and fresh Ph.D. Students might be inspired by our work and, further contributing to the field. All in all, our work brings the following major contributions:

\begin{enumerate}
    \item A systematic literature review on non-functional attributes of machine learning-intensive systems, which allowed to systematically classify them, other than identify the current open challenges to cope with them. These pieces of information may be exploited by researchers to define the next research steps to improve the currently available techniques and overall support provided to practitioners;

    \smallskip
    \item A set of implications and take-away messages that researchers may use to address future research avenues on the management and optimization of non-functional requirements of machine learning-enabled systems;

    \smallskip
    \item An online appendix containing all data and scripts used in the study, which might be used to extend the scope of the study or replicate our work.
    
\end{enumerate}

The considerations and implications of this systematic literature review drive our future research agenda. We first aim at deepening our analysis of how the research community has investigated non-functional requirements so far, possibly looking for scientific papers that, despite not explicitly targeting non-functional requirements, tangentially improve non-functional properties of machine learning-enabled systems. We will also work toward the definition of novel approaches that could optimize the non-functional requirements of machine learning-enabled systems while keeping accuracy into account. In addition, we aim to design software analytics studies, surveys and semi-structured interviews to further understand the impact of non-functional requirements and the practitioners' expectations with respect to them.